\documentclass[twocolumn]{aastex7}

\usepackage[flushleft]{threeparttable}

\usepackage{xspace}
\usepackage{array}
\usepackage{makecell}

\newcommand{\msolpcsq}{M$_\odot$\,pc$^{-2}$\xspace}

\newcommand{\msol}{M$_\odot$\xspace}

\newcommand{\hi}{\ion{H}{1}\xspace}

\newcommand{\hii}{{\sc H\textsc{ ii}}\xspace}
\newcommand{\kms}{km s$^{-1}$\xspace}

\def\lesssim{\mathrel{\hbox{\rlap{\hbox{\lower4pt\hbox{$\sim$}}}\hbox{$<$}}}}
\def\gtrsim{\mathrel{\hbox{\rlap{\hbox{\lower4pt\hbox{$\sim$}}}\hbox{$>$}}}}

\shorttitle{The Local Group L Band Survey}
\shortauthors{LGLBS Collaboration}

\graphicspath{{./}{figures/}}

\begin{document}

\title{The Karl G. Jansky Very Large Array Local Group L-band Survey (LGLBS)}

\newcommand{\Ox}{Sub-department of Astrophysics, Department of Physics, University of Oxford, Keble Road, Oxford OX1 3RH, UK}

\newcommand{\UGent}{Sterrenkundig Observatorium, Universiteit Gent, Krijgslaan 281 S9, B-9000 Gent, Belgium}

\newcommand{\STScI}{Space Telescope Science Institute, 3700 San Martin Drive, Baltimore, MD 21218, USA}

\newcommand{\IWR}{Universit\"{a}t Heidelberg, Interdisziplin\"{a}res Zentrum f\"{u}r Wissenschaftliches Rechnen, Im Neuenheimer Feld 205, 69120 Heidelberg, Germany}
\newcommand{\Radcliffe}{Elizabeth S. and Richard M. Cashin Fellow at the Radcliffe Institute for Advanced Studies at Harvard University, 10 Garden Street, Cambridge, MA 02138, U.S.A.}

\newcommand{\MPIA}{Max-Planck-Institut f\"{u}r Astronomie, K\"{o}nigstuhl 17, D-69117, Heidelberg, Germany}

\newcommand{\AURA}{AURA for the European Space Agency (ESA), Space Telescope Science Institute, 3700 San Martin Drive, Baltimore, MD 21218, USA}

\newcommand{\UCSD}{Department of Astronomy \& Astrophysics, University of California, San Diego, 9500 Gilman Dr., La Jolla, CA 92093, USA}

\newcommand{\Whitman}{Whitman College, 345 Boyer Avenue, Walla Walla, WA 99362, USA}

\newcommand{\JHU}{Department of Physics and Astronomy, The Johns Hopkins University, Baltimore, MD 21218, USA}

\newcommand{\OSU}{Department of Astronomy, The Ohio State University, 140 West 18th Avenue, Columbus, OH 43210, USA}

\newcommand{\OSUPhys}{Department of Physics, The Ohio State University, Columbus, Ohio 43210, USA}

\newcommand{\CCAPP}{Center for Cosmology and Astroparticle Physics (CCAPP), 191 West Woodruff Avenue, Columbus, OH 43210, USA}

\newcommand{\ARI}{Astronomisches Rechen-Institut, Zentrum f\"{u}r Astronomie der Universit\"{a}t Heidelberg, M\"{o}nchhofstr. 12-14, D-69120 Heidelberg, Germany}

\newcommand{\ANU}{Research School of Astronomy and Astrophysics, Australian National University, Canberra, ACT 2611, Australia}

\newcommand{\UConn}{Department of Physics, University of Connecticut, 196A Auditorium Road, Storrs, CT 06269, USA}

\newcommand{\UHawaii}{Institute for Astronomy, University of Hawaii, 2680 Woodlawn Drive, Honolulu, HI 96822, USA}

\newcommand{\UniCA}{Universit\'{e} C\^{o}te d'Azur, Observatoire de la C\^{o}te d'Azur, CNRS, Laboratoire Lagrange, 06000, Nice, France}

\newcommand{\UAlberta}{Dept. of Physics, University of Alberta, 4-183 CCIS, Edmonton, Alberta, T6G 2E1, Canada}

\newcommand{\Arcetri}{INAF — Osservatorio Astrofisico di Arcetri, Largo E. Fermi 5, I-50125, Florence, Italy}

\newcommand{\UWyoming}{Department of Physics and Astronomy, University of Wyoming, Laramie, WY 82071, USA}

\newcommand{\LJMU}{Astrophysics Research Institute, Liverpool John Moores University, 146 Brownlow Hill, Liverpool L3 5RF, UK}

\newcommand{\ITA}{Universit\"{a}t Heidelberg, Zentrum f\"{u}r Astronomie, Institut f\"{u}r Theoretische Astrophysik, Albert-Ueberle-Str 2, D-69120 Heidelberg, Germany}

\newcommand{\CfA}{Center for Astrophysics $\mid$ Harvard \& Smithsonian, 60 Garden St., 02138 Cambridge, MA, USA}

\newcommand{\MPE}{Max-Planck-Institut f\"{u}r Extraterrestrische Physik (MPE), Giessenbachstr. 1, D-85748 Garching, Germany}

\newcommand{\UMD}{Department of Astronomy and Joint Space-Science Institute, University of Maryland, College Park, MD 20742, USA}

\newcommand{\UVA}{Department of Astronomy, University of Virginia, Charlottesville, VA, USA}

\newcommand{\NRAO}{National Radio Astronomy Observatory, Charlottesville, VA, USA}

\newcommand{\ASIAA}{Institute of Astronomy and Astrophysics, Academia Sinica, No. 1, Sec. 4, Roosevelt Road, Taipei 106216, Taiwan}

\newcommand{\kipac}{Kavli Institute for Particle Astrophysics \& Cosmology (KIPAC), Stanford University, CA 94305, USA}

\newcommand{\aifa}{Argelander-Institut f\"ur Astronomie, University of Bonn, Auf dem H\"ugel 71, 53121 Bonn, Germany}

\newcommand{\TKU}{Department of Physics, Tamkang University, No.151, Yingzhuan Road, Tamsui District, New Taipei City 251301, Taiwan}

\newcommand{\CarnegieObs}{The Observatories of the Carnegie Institution for Science. 813 Santa Barbara Street, Pasadena, CA 91101, USA}

\newcommand{\Princeton}{Department of Astrophysical Sciences, Princeton University, Princeton, NJ 08544, USA}

\newcommand{\IAS}{Institute for Advanced Study, 1 Einstein Drive, Princeton, NJ 08540, USA}

\newcommand{\COOL}{Cosmic Origins Of Life (COOL) Research DAO, coolresearch.io}

\newcommand{\ESO}{European Southern Observatory (ESO), Karl-Schwarzschild-Stra{\ss}e 2, 85748 Garching, Germany}

\newcommand{\ULyon}{Univ Lyon, Univ Lyon 1, ENS de Lyon, CNRS, Centre de Recherche Astrophysique de Lyon UMR5574, F-69230 Saint-Genis-Laval, France}

\newcommand{\UoM}{Jodrell Bank Centre for Astrophysics, Department of Physics and Astronomy, University of Manchester, Oxford Road, Manchester M13 9PL, UK}

\correspondingauthor{Eric W. Koch}
\email{koch.eric.w@gmail.com}

\author[0000-0001-9605-780X]{Eric W. Koch}
\affiliation{National Radio Astronomy Observatory, 800 Bradbury SE, Suite 235, Albuquerque, NM 87106}
\affiliation{\CfA}
\email{koch.eric.w@gmail.com}

\author[0000-0002-2545-1700]{Adam~K.~Leroy}
\affiliation{\OSU}
\affiliation{\CCAPP}
\email{leroy.42@osu.edu}

\author[0000-0002-5204-2259]{Erik~W.~Rosolowsky}
\affiliation{\UAlberta}
\email{rosolowsky@ualberta.ca}

\author[0000-0002-8400-3705]{Laura Chomiuk}
\affiliation{Center for Data Intensive and Time Domain Astronomy, Department of Physics and Astronomy,\\ Michigan State University, East Lansing, MI 48824, USA}
\email{chomiukl@msu.edu}

\author[0000-0002-1264-2006]{Julianne J.~Dalcanton}
\affiliation{Center for Computational Astrophysics, Flatiron Institute, 162 Fifth Avenue, New York, NY 10010, USA}
\affiliation{Department of Astronomy, Box 351580, University of Washington, Seattle, WA 98195, USA}
\email{jdalcanton@flatironinstitute.org}

\author[0000-0001-9504-7386]{Nickolas~M.~Pingel}
\affiliation{University of Wisconsin–Madison, Department of Astronomy, 475 N Charter St, Madison, WI 53703, USA}
\email{nmpingel@wisc.edu}

\author[0000-0002-2545-1700]{Sumit K. Sarbadhicary}
\affiliation{\OSU}
\affiliation{\CCAPP}
\affiliation{\JHU}
\email{ssarbad1@jh.edu}

\author[0000-0002-3418-7817]{Sne{\v z}ana Stanimirovi{\'c}}
\affiliation{University of Wisconsin–Madison, Department of Astronomy, 475 N Charter St, Madison, WI 53703, USA}
\email{sstanimi@astro.wisc.edu}

\author[0000-0003-4793-7880]{Fabian Walter}
\affiliation{Max-Planck-Institut f\"ur Astronomie, K\"onigstuhl 17, 69117 Heidelberg, Germany
}
\email{walter@mpia.de}

\author[0000-0002-8449-4815]{Haylee N. Archer} \affiliation{School of Earth and Space Exploration, Arizona State University, Tempe, AZ 85287, USA} \affiliation{Lowell Observatory, 1400 W Mars Hill Rd, Flagstaff, AZ 86001, USA}
\email{harcher@lowell.edu}

\author[0000-0002-5480-5686]{Alberto D. Bolatto}\affiliation{Department of Astronomy and Joint Space-Science Institute, University of Maryland, College Park, MD 20854}
\email{bolatto@umd.edu}

\author[0000-0003-4961-6511]{Michael P. Busch}
\affiliation{Department of Astronomy \& Astrophysics, University of California, San Diego, 9500 Gilman Drive, La Jolla, CA 92093, USA}
\email{mpbusch@ucsd.edu}

\author[0009-0005-1781-5665]{Hongxing Chen}
\affiliation{University of Wisconsin–Madison, Department of Astronomy, 475 N Charter St, Madison, WI 53703, USA}
\email{hchen792@wisc.edu}

\author[0000-0001-8241-7704]{Ryan Chown}\affiliation{Department of Astronomy, The Ohio State University, 140 West 18th Avenue, Columbus, OH 43210, USA}
\email{chown.5@osu.edu}

\author[0009-0007-7949-6633]{Harrisen Corbould}\affiliation{\UAlberta}
\email{harrisen@ualberta.ca}

\author[0000-0002-9511-1330]{Serena A. Cronin}\affiliation{Department of Astronomy, University of Maryland, College Park, MD 20854}
\email{cronin@umd.edu}

\author[0000-0003-2511-2060]{Jeremy Darling}
\affiliation{Center for Astrophysics and Space Astronomy, Department of Astrophysical and Planetary Sciences, University of Colorado, 389 UCB, Boulder, CO 80309, USA}
\email{jeremy.darling@colorado.edu}

\author[0000-0001-6105-7308]{Thomas Do}
\affiliation{Center for Data Intensive and Time Domain Astronomy, Department of Physics and Astronomy,\\ Michigan State University, East Lansing, MI 48824, USA}
\email{dothomas@msu.edu}

\author[0000-0002-3106-7676]{Jennifer Donovan Meyer}\affiliation{National Radio Astronomy Observatory, 520 Edgemont Road, Charlottesville, VA 22903, USA}
\email{jmeyer@nrao.edu}

\author[0000-0002-1185-2810]{Cosima Eibensteiner}\altaffiliation{Jansky Fellow of the National Radio Astronomy Observatory}\affiliation{National Radio Astronomy Observatory, 520 Edgemont Road, Charlottesville, VA 22903, USA}
\email{ceibenst@nrao.edu}

\author[0000-0002-3322-9798]{Deidre Hunter}\affiliation{Lowell Observatory, Flagstaff, AZ, 86001, USA}
\email{dah@lowell.edu}

\author[0000-0002-4663-6827] {R\'emy Indebetouw} \affiliation{University of Virginia Astronomy Department, P.O. Box 400325, Charlottesville, VA, 22904, USA}
\affiliation{National Radio Astronomy Observatory, 520 Edgemont Rd, Charlottesville, VA 22903, USA}
\email{rindebet@nrao.edu}

\author[0000-0002-5825-9635]{Preshanth Jagannathan} \affiliation {National Radio Astronomy Observatory, 1011 Lopezville Road, Socorro, NM 87801, USA}
\email{pjaganna@nrao.edu}

\author[0000-0002-3227-4917]{Amanda A. Kepley} \affiliation{National Radio Astronomy Observatory, 520 Edgemont Road, Charlottesville, VA 22903, USA}
\email{akepley@nrao.edu}

\author[0000-0003-2896-3725]{Chang-Goo Kim}
\affiliation{Department of Astrophysical Sciences, Princeton University, 4 Ivy Lane, Princeton, NJ 08544, USA}
\email{changgoo@princeton.edu}

\author[0000-0002-6760-7531]{Shin-Jeong Kim}
\affiliation{Research School of Astronomy and Astrophysics, The Australian National University, Canberra ACT 2611, Australia}
\email{Shin-Jeong.Kim@anu.edu.au}

\author[0000-0001-6649-8559]{Timea O. Kovacs}
\affiliation{\MPIA}
\email{kovacs@mpia.de}

\author[0000-0003-1111-8066]{Joshua Marvil}
\affiliation{National Radio Astronomy Observatory, 1011 Lopezville Road, Socorro, NM 87801, USA}
\email{jmarvil@nrao.edu}

\author[0000-0001-7089-7325]{Eric J. Murphy}
\affiliation{National Radio Astronomy Observatory, 520 Edgemont Road, Charlottesville, VA 22903, USA}
\email{emurphy@nrao.edu}

\author[0000-0002-7743-8129]{Claire E. Murray}
\affil{Space Telescope Science Institute, 3700 San Martin Drive, Baltimore, MD 21218, USA}
\affil{The William H. Miller III Department of Physics \& Astronomy, Bloomberg Center for Physics and Astronomy, Johns Hopkins University, 3400 N. Charles Street, Baltimore, MD 21218, USA}
\email{clairemurray56@gmail.com}

\author[0000-0001-8224-1956]{J\"urgen Ott}
\affiliation{National Radio Astronomy Observatory, 1011 Lopezville Road, Socorro, NM 87801, USA}
\email{jott@nrao.edu}

\author[0000-0001-7996-7860]{D.J. Pisano}
\affiliation{Dept. of Astronomy, Univ. of Cape Town, Private Bag X3, Rondebosch 7701, South Africa}
\email{pisano@ast.uct.ac.za}

\author[0000-0002-1129-1873]{Mary Putman} \affiliation{Department of Astronomy, Columbia University, New York, NY 10027, USA}
\email{mputman@astro.columbia.edu}

\author[0000-0003-3351-6831]{Daniel R. Rybarczyk}
\affiliation{University of Wisconsin--Madison, Department of Astronomy, 475 N Charter St, Madison, WI 53703, USA}
\email{rybarczyk@astro.wisc.edu}

\author[0000-0001-6326-7069]{Julia Roman-Duval}
\affiliation{Space Telescope Science Institute, 3700 San Martin Drive, Baltimore, MD 21218, USA}
\email{duval@stsci.edu}

\author[0000-0002-4378-8534]{Karin~Sandstrom} \affiliation{Department of Astronomy \& Astrophysics, University of California, San Diego, 9500 Gilman Dr., La Jolla, CA 92093, USA}
\email{karin.sandstrom@gmail.com}

\author[0000-0002-3933-7677]{Eva Schinnerer}
\affiliation{\MPIA}
\email{schinner@mpia.de}

\author[0000-0003-0605-8732]{Evan D. Skillman}
\affiliation{Minnesota Institute for Astrophysics, University of Minnesota, Minneapolis, MN 55455, USA}
\email{skill001@umn.edu}

\author[0000-0003-2599-7524]{Adam Smercina}\thanks{Hubble Fellow}
\affiliation{Space Telescope Science Institute, 3700 San Martin Dr., Baltimore, MD 21218, USA}
\email{asmerci@uw.edu}

\author[0000-0002-4110-8769]{Ioana Stelea}
\affiliation{University of Wisconsin–Madison, Department of Astronomy, 475 N Charter St, Madison, WI 53703, USA}
\email{stelea@wisc.edu}

\author[0000-0002-1468-9668]{Jay Strader}
\affiliation{Center for Data Intensive and Time Domain Astronomy, Department of Physics and Astronomy,\\ Michigan State University, East Lansing, MI 48824, USA}
\email{straderj@msu.edu}

\author[0000-0003-0378-4667]{Jiayi~Sun}\thanks{Hubble Fellow}
\affiliation{Department of Astrophysical Sciences, Princeton University, 4 Ivy Lane, Princeton, NJ 08544, USA}
\email{jiayi.sun@princeton.edu}

\author[0009-0000-2209-7972]{Devisree Tallapaneni}
\affiliation{\OSU}
\email{devisree.tallapaneni@gmail.com}

\author[0000-0003-1356-1096]{Elizabeth Tarantino} \affil{Space Telescope Science Institute, 3700 San Martin Drive, Baltimore, MD 21218, USA}
\email{etarantino@stsci.edu}

\author[0000-0002-5877-379X]{Vicente Villanueva}
\affiliation{Departamento de Astronom\'ia, Universidad de Concepci\'on, Barrio Universitario, Concepci\'on, Chile}
\email{vvillanueva@astro-udec.cl}

\author[0000-0002-6442-6030]{Daniel R. Weisz} 
\affiliation{Department of Astronomy, University of California, Berkeley, Berkeley, CA, 94720, USA}
\email{dan.weisz@berkeley.edu}

\author[0000-0002-0012-2142]{Thomas~G.~Williams}
\affiliation{Subdepartment of Astrophysics, Department of Physics, University of Oxford, Keble Road, Oxford OX1 3RH, UK}
\email{thomas.williams@physics.ox.ac.uk}

\author[0000-0002-7759-0585]{Tony Wong}
\affiliation{Department of Astronomy, University of Illinois, Urbana, IL 61801, USA}
\email{wongt@illinois.edu}

\begin{abstract}
We present the Local Group L-Band Survey (LGLBS), a Karl G. Jansky Very Large Array (VLA) survey producing the highest quality 21-cm and $1{-}2$~GHz radio continuum images to date for the six VLA-accessible, star-forming, Local Group galaxies. Leveraging the VLA's spectral multiplexing power, we simultaneously survey the 21-cm line at high $0.4$ km s$^{-1}$ velocity resolution, the 1--2 GHz polarized continuum, and four OH lines. For the massive spiral M31, the dwarf spiral M33, and the dwarf irregular galaxies NGC6822, IC10, IC1613, and the Wolf-Lundmark-Melotte Galaxy (WLM), we use all four VLA configurations and the Green Bank Telescope to reach angular resolutions of $< 5''$ ($10{-}20$~pc) for the 21-cm line with $<10^{20}$~cm$^{-2}$ column density sensitivity, and even sharper views ($< 2''$; $5{-}10$~pc) of the continuum. Targeting these nearby galaxies ($D\lesssim1$ Mpc) reveals a sharp, resolved view of the atomic gas, including 21-cm absorption, and continuum emission from supernova remnants and \ion{H}{2} regions. These datasets can be used to test theories of the abundance and formation of cold clouds, the driving and dissipation of interstellar turbulence, and the impact of feedback from massive stars and supernovae. Here, we describe the survey design and execution, scientific motivation, data processing, and quality assurance. We provide a first look at and publicly release the wide-field 21-cm \hi\ data products for M31, M33, and four dwarf irregular targets in the survey, which represent some of the highest physical resolution 21-cm observations of any external galaxies beyond the LMC and SMC.
\end{abstract}


\section{Introduction}
\label{sec:intro}

Atomic gas (\ion{H}{1}) makes up most of the interstellar medium (ISM) in most galaxies at all redshifts, and represents a significant majority of the gas in the disks of galaxies at $z=0$ \citep{Tacconi2020-highzreview,Walter2020,Saintonge2022}. The structure of this atomic gas on scales of $\approx 10{-}100$~pc is shaped by turbulence, stellar feedback, and the combination of thermal and dynamical instabilities that govern the formation and destruction of cold clouds \citep{Dobbs2014,McClureGriffiths2023}. Meanwhile, the \ion{H}{2} regions and supernova remnants (SNRs) represent the two most important sources of stellar feedback sources: young, massive stars and supernova explosions \citep{Schinnerer2024}.

The L-Band ($\nu = 1{-}2$~GHz) contains key diagnostics of atomic gas, \ion{H}{2} regions, and SNRs. The 21-cm hyperfine line of hydrogen \citep[$\nu = 1420$~MHz,][]{Vanderhulst1945,EwenPurcell1951} traces the mass, motions, and (combining extinction and absorption) temperature of the atomic ISM \citep{Kerr1954,field1958}. Meanwhile, the $1{-}2$~GHz radio emission suffers no extinction due to dust, and so offers a direct view of \ion{H}{2} regions via free-free emission and SNRs via synchrotron emission, capturing key stages of the life and death of massive stars. Beyond these tracers, the L-band also hosts a suite of other, fainter features that access other aspects of the matter cycle. These include diffuse synchrotron and free-free emission, molecular OH transitions at 1612, 1665, 1667, and 1712 MHz, radio recombination lines, and polarization of the continuum.

Unfortunately, the relatively coarse angular resolution of current radio interferometers when configured to achieve the required surface brightness sensitivity has limited our ability to utilize these diagnostics at high physical resolution. Current facilities can observe 21-cm emission at $\sim 5''$ and $1{-}2$~GHz continuum at $\sim 2''$. This is significantly coarser than the $0.1{-}1''$ regularly achieved by facilities studying other phases the ISM, including \textit{ALMA} and \textit{JWST}. This situation is expected to change with the construction of a next-generation Very Large Array \citep[ngVLA; ][]{Murphy2018} and the Square Kilometre Array \citep{SKA}. Until then, matching the $\gtrsim 2{-}5''$ angular scales accessible to L-band radio facilities to the small physical scales of individual gas clouds, SNRs, or \ion{H}{2} regions requires targeting the nearest galaxies.

Galaxies in the Local Group, with $D < 1.2$~Mpc \citep[][]{vandenbergh1999}, represent key targets that maximize the physical resolution and mass or luminosity sensitivity of any observations. These galaxies thus offer the best opportunity to detect and resolve individual gas clouds, SNRs, and \ion{H}{2} regions, in full context. Because of their proximity, these targets tend to be observed in detail by almost all astronomical facilities, and have seen major investment by facilities from \textit{HST} \citep[e.g.,][]{Dalcanton2012,Williams2021} and \textit{ALMA} \citep[e.g.,][]{Rubio2015,Schruba2017} to \textit{Euclid} \citep[e.g.,][]{hunt2024} and \emph{JWST} \citep[e.g.,][]{Peltonen2024,Jones2023,Weisz2024}.

Motivated by these considerations, we present the Local Group L-Band Survey (LGLBS), a Karl G.\ Jansky Very Large Array (VLA) survey designed to address open questions related to the physics of atomic gas, star formation, cosmic rays, and stellar feedback, and to provide a suite of legacy data products. LGLBS targets the six northern, VLA-accessible star-forming Local Group galaxies using all four of the VLA's main configurations (A, B, C, D, from most extended to compact), and includes short-spacing corrections from the Robert C. Byrd Green Bank Telescope for the atomic gas maps. Fortunately, the VLA-accessible Local Group galaxies targeted by LGLBS span a diverse set of physical conditions. M31 (Andromeda) offers us an external view of a close cousin to the Milky Way, M33 (Triangulum) often acts as the prototype dwarf spiral galaxy, and four actively star-forming dwarf irregular galaxies provide windows into the common phenomenon of \ion{H}{1} dominated, star-forming, low metallicity dwarf galaxies. 

The LGLBS project is intended as both a high-detail successor to the successful VLA 21-cm surveys THINGS, LITTLE THINGS, and VLA-ANGST \citep{Walter2008,Hunter2012,Ott2012}, as well as a preview of the type of observations that the ngVLA and SKA will be able to produce for more distant targets. Most importantly, the project aims to produce sharp views of the atomic gas and radio continuum that develop our understanding of several important but less-studied aspects of the matter cycle in galaxies. We also aim to provide widely useful public data products for this well-studied set of nearest galaxies.

This paper describes the LGLBS science drivers (Section \ref{sec:science_drivers}), targets (Section \ref{sec:lg_sample}), observing strategy (Section \ref{sec:survey}), and data reduction and quality assurance procedures (Section \ref{sec:dr_qa}). As a demonstration of the results, we present 21-cm \ion{H}{1} imaging using the VLA C and D configurations (hereafter CD) (Section \ref{sec:lglbs_hi_cd}). Companion papers will detail the full A, B, C, and D (ABCD) configuration 21-cm imaging (Pingel et al. in preparation) and radio continuum imaging (Sarbadhicary et al. in preparation).

\section{Scientific Motivation}
\label{sec:science_drivers}

At $z=0$, atomic hydrogen (\hi) makes up the majority of neutral gas in the universe, with $\sim 3$ times more \hi\ than H$_2$ in galaxies \citep{Carilli2013,Saintonge2022}. \hi\ represents the material for molecule and cloud formation, and the ability to convert atomic to molecular gas regulates star formation in many cases \citep[][]{Leroy2008,Hunter2024}. The \hi\ also bears the brunt of much of the stellar feedback in galaxies, from photoelectric heating to supernova feedback. This leads to important phenomena including fountains, outflows, the carving of shells and holes, and the driving of interstellar turbulence \citep[][]{Elmegreen2002,Elmegreen2004,MacLow2004,Sancisi2008, Fraternali2017,Thompson2024}. These phenomena affect and are driven by the physical state of the gas, and atomic gas spans a wide range of physical conditions, with densities from $\sim 0.1{-}100$~cm$^{-3}$ and temperatures from $\sim10^2-10^4$~K \citep[e.g.,][]{McClureGriffiths2023}. Unlike the mass of atomic gas, the physical state of the \hi\ is not readily accessible from low resolution 21-cm observations.

LGLBS aims to resolve the 21-cm emission tracing \hi\ down to the scales of individual star-forming regions and ISM clouds ($\sim 25$ pc) to diagnose the physical properties of the atomic gas and measure its dynamical and morphological state. Measuring the \hi\ associated with individual molecular-atomic complexes will provide sharp tests of current models used to explain the molecular-to-atomic gas ratio and the rate of star formation in galaxies \citep[e.g.,][]{Krumholz2014,Sternberg2014,Wolfire2022}. LGLBS will also allow identification of atomic gas filaments, bound structures, outflows, shells,  and extraplanar atomic gas clouds analogous to the high velocity clouds seen in the Milky Way. In the Milky Way, such observations have proven critical to assess the role of magnetic fields and turbulence, assess the importance of various instabilities, and study the mechanisms for gas accretion onto the galaxy \citep[see][]{putman2012,Dobbs2014,McClureGriffiths2023,Mittal2023,Hacar2023}. However, so far this level of detail has been mostly inaccessible in galaxies beyond the Magellanic Clouds. 

In addition to high physical resolution, the LGLBS velocity resolution, $0.4$~km~s$^{-1}$, is sufficient to resolve thermal line widths for cold $\lesssim 100$~K atomic gas. Further, the large areal coverage of the observations and the high angular and velocity resolution allow searches for 21-cm \emph{absorption}, not only emission. Paired emission and absorption measurements are a crucial tool for measuring the spin temperature and opacity of the 21-cm line \citep[e.g.,][]{Heiles2003,Murray2014,Murray2018}. Demonstrating this capability, LGLBS has already yielded the first detection of 21-cm absorption in a low metallicity galaxy (NGC6822) beyond the Magellanic Clouds \citep{Pingel2024}.

The inclusion of all VLA configurations and short-spacing GBT data allows LGLBS to recover structure across all scales. As result LGLBS will be ideal to study the nature and drivers of interstellar turbulence \citep[e.g.,][]{Koch2020}. The survey meshes naturally with extensive previous investment in these targets. These include massive efforts by \emph{HST} to map the stellar populations and so capture the sources of stellar feedback \citep[][]{holtzman2006, Dalcanton2012,Williams2021}, efforts by \textit{Spitzer} and \textit{Herschel} to map the dust \citep[e.g.,][]{Draine2014,RemyRuyer2014}, and multiple facilities, including \textit{ALMA}, to observe the molecular gas via the CO lines \citep[e.g.,][]{ohta1993,Leroy2006,Nieten2006,Rubio2015,Schruba2017,Muraoka2023}.

In addition to capturing background sources useful for 21-cm absorption studies, the 1--2 GHz radio continuum offers a new view of \hii\ regions and SNRs, tracing recent massive star formation and stellar feedback. Emission in this frequency range is a mixture of free-free and synchrotron emission. The free-free emission provides an extinction-free tracer of the ionizing photon production rate and a probe of \hii\ region structure analogous to hydrogen recombination lines. In the Local Group, our multi-configuration continuum imaging corresponds to $\sim 5{-}10$~pc resolution, sufficient to resolve most evolved \hii\ regions \citep[e.g.,][]{Anderson2014}. Based on the star formation rates of our targets, we expect to measure the free-free emission associated with $\gtrsim 5,000$ optically-visible \hii\ regions \citep[e.g.,][]{Azimlu2011}. Because the free-free emission is unaffected by extinction, LGLBS also offers the opportunity to discover new, embedded regions missed by optical surveys.

The 1--2~GHz continuum also traces supernova remnants through the synchrotron radiation from shock waves interacting with the ISM, amplifying magnetic fields, and accelerating relativistic particles \citep[e.g., ][]{Chomiuk2009,White2019}. Radio synchrotron emission from SNRs remains visible for 20--80 kyr, i.e., throughout the ejecta-dominated and Sedov-Taylor phases \citep[e.g.,][]{Berezhko2004}. Unlike optical and X-ray diagnostics of SNRs (e.g., \citealt{Lee2014, Sasaki2012}), radio emission is unaffected by absorption from dust and gas. On more extended scales, the 1--2 GHz synchrotron emission traces cosmic ray electrons accelerated in SN shocks and diffusing across the galaxy, which acts on large scales as a star formation tracer \citep[e.g.,][]{Condon1992} but with additional dependence on the magnetic field, ISM density, and aging of the cosmic ray $e^{-}$ population.

Beyond mapping the atomic gas, \hii\ regions, and SNRs, the LGLBS spectral setup also captures the polarized radio continuum and line emission from the OH molecule.  These tracers are fainter but offer potentially important constraints on ISM conditions. The polarized radio continuum traces the geometry of the magnetic fields and constrains field strength \citep{Beck2015}. The OH transitions at 1.612, 1.665, 1.667, and 1.712 GHz offer the prospect to find new OH masers or trace molecular gas, including CO-dark gas, in emission or absorption \citep[e.g.,][]{Allen1995, Busch2024}. 

These goals mean that LGLBS represents a natural complement to the new GASKAP-HI \citep[][]{Pingel2022} and GASKAP-OH \citep[][]{Brown2024} surveys of the Magellanic Clouds and Milky Way and recent investment in emission, absorption, and continuum studies of the Milky Way plane \citep[e.g., GALFA, 21-SPONGE, THOR, and the MeerKAT Galactic Plane Survey,][]{Peek2011,Murray2014,beuther2016,Murray2018,Padmanabh2023}. Our goal is to expand a similarly detailed view to an external perspective, covering the diverse northern Local Group galaxies.

\section{The Local Group Sample}
\label{sec:lg_sample}

\begin{figure*}[!t]
    \centering
    \includegraphics[width=1.0\textwidth]{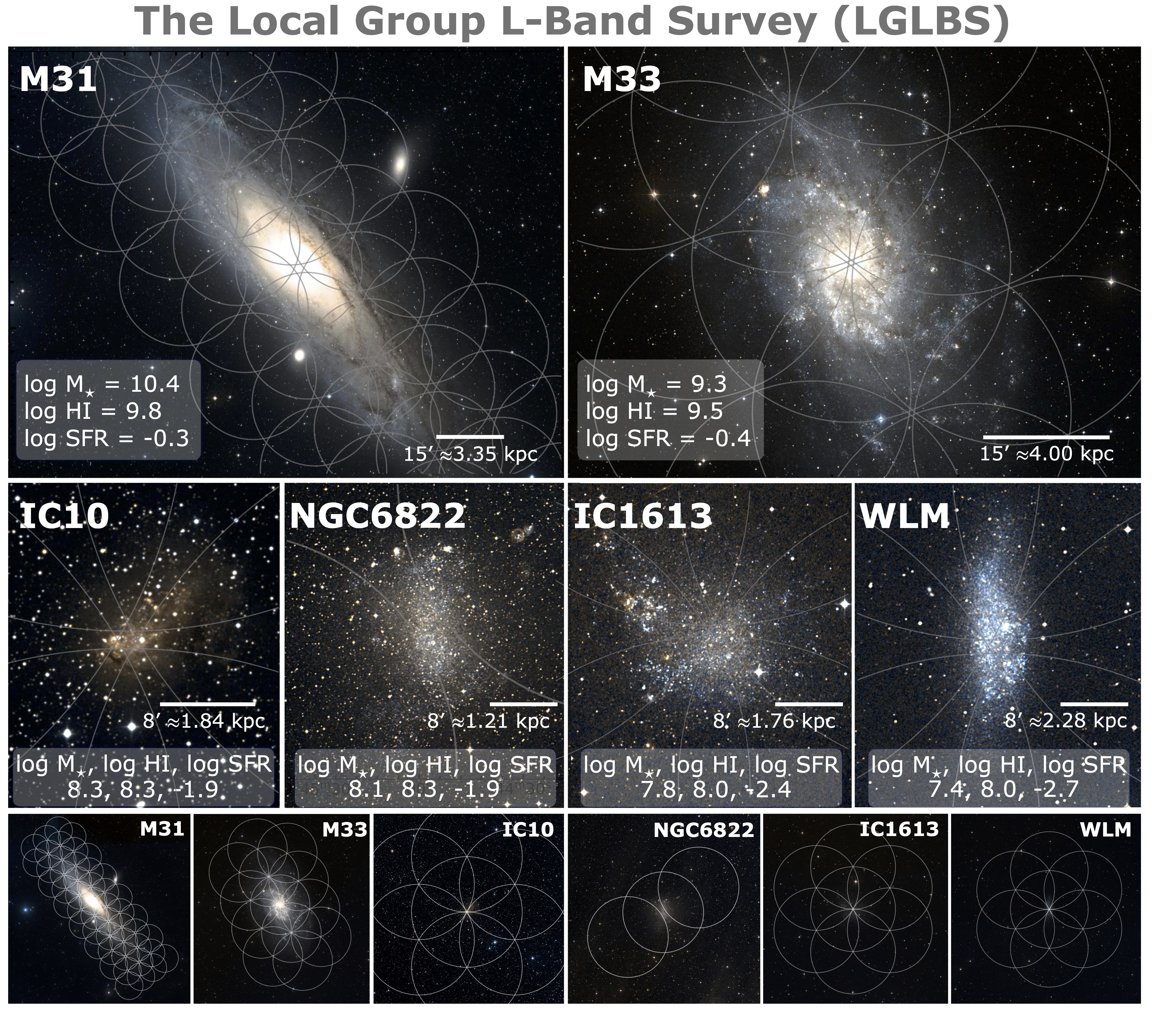}
    \caption{Three-color optical images showing the six northern hemisphere local galaxies ordered by decreasing stellar mass (image credits: DESI Legacy Survey, \citealt{desi1,desi2,desi3}) overlaid with the VLA pointings. The scale bars on the lower right corners indicate about one radius (for M31 and M33) and a half of the radius (for IC10, NGC6822, IC1613, and WLM) of one VLA pointing. The lower row shows the full extent of the LGLBS pointings.}
    \label{fig:optical_collage}
\end{figure*}

There are eight Local Group \citep[$D < 1.2$~Mpc][]{vandenbergh1999} galaxies with on-going star formation (i.e., UV or H$\alpha$ emission) and that are gas rich \citep[$M_{\rm HI} \gtrsim 10^{7.5}$~M$_\odot$;][]{Karachentsev2004,Karachentsev2013}. LGLBS observes the six of these targets visible to the VLA (declination $\gtrsim -35^{\circ}$; Figure \ref{fig:optical_collage}). These are the massive spiral M31, the dwarf spiral M33, and four dwarf irregulars: IC10, IC1613, NGC6822, and WLM. The other two galaxies with gas masses $>10^8~M_\odot$ are the LMC and SMC, which are located in the Southern hemisphere and currently being observed by SKA precursors \citep[e.g., GASKAP;][]{GASKAP,Pingel2022}. 

\begin{figure*}[!t]
\centering
\includegraphics[width=0.9\textwidth]{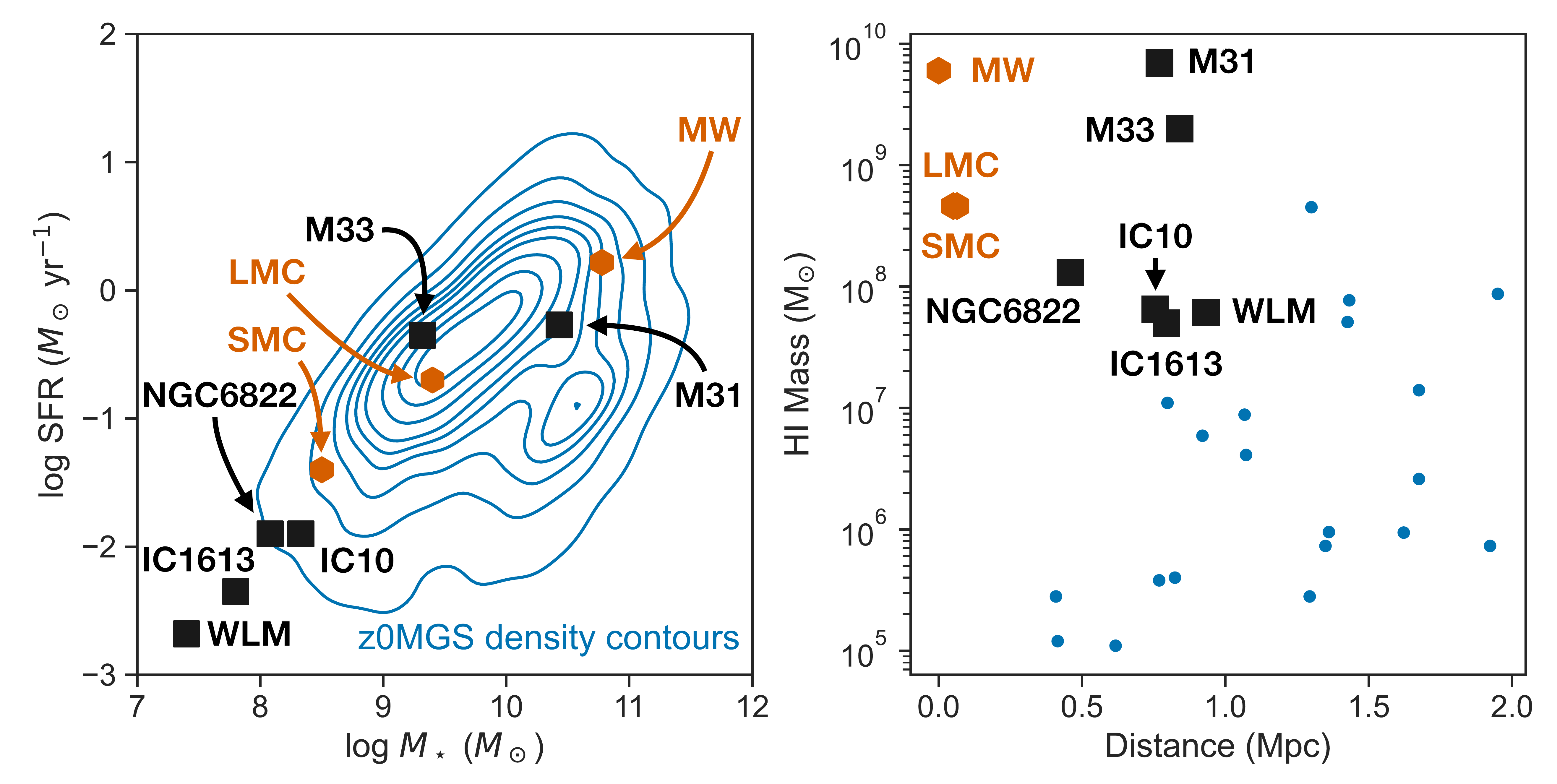}
\caption{Left: Star formation rate versus stellar mass of the LGLBS targets, the SMC, LMC, and Milky Way relative to the local galaxy population ($d<50$~Mpc) from z0MGS \citep[an archival project combining WISE and GALEX images of nearby galaxies that is complete to ${\rm M}_\star > 10^9$~\msol;][]{Leroy2019}.
Right: \hi\ mass versus distance for Local Group galaxies  with \hi\ detections from the sample compiled in \citet{Putman2021}. 
The LGLBS \hi\ masses are given in Table \ref{tab:physical} from this work.
Milky Way values adopted from \citet{Licquia2015} and \citet{Ferriere2001}.
}
\label{fig:lglbs_sample_comparison}
\end{figure*}

Although it remains a small sample, the Local Group captures a large diversity in the conditions relevant to ISM physics, star formation, and galaxy evolution. We summarize the properties of our targets in Table \ref{tab:physical}, and Figure \ref{fig:lglbs_sample_comparison} shows these properties in the broader context of the local galaxy population. M31 is a massive, dust- and metal-rich spiral, with low-intensity star formation, an extended \hi\ disk, and a flat rotation curve \citep{Corbelli2010}. While similar in scale to the Milky Way, M31 provides an external perspective on the ISM, star formation, and galactic structure.  M33 is a dwarf spiral, comparable in mass and star formation properties to the LMC, but it features a comparatively undisturbed flocculent disk \citep{Corbelli2003}. The four low-mass dwarf galaxies (IC10, IC1613, NGC6822, WLM) have irregular ISM morphologies dominated by extended \hi\ distributions and giant shells, which are driven by stellar feedback \citep[e.g.,][see Table \ref{tab:lit} for more references]{Wilcots1998,deBlok2006,Hunter2012}. Compared to the Milky Way and M31, these low-mass dwarf galaxies also have lower dust and metal abundances, by factors of 3-10, solid body or dispersion-dominated kinematics, and extended \hi\ reservoirs \citep{Mateo1998,Chiang2021}. Our targets overlap the LMC and SMC in stellar mass and metallicity but differ in key ways. Most notably, the recent star formation history and large-scale ISM structure in the LMC and SMC are heavily impacted by on-going interactions with each other and with the Milky Way halo \citep{Stanimirovic1999,Putman2003,Besla2012,Joshi2019,Lucchini2021} while the LGLBS targets are more isolated \citep{Mateo1998}. The combination of the four LGLBS dwarfs with the SMC also offers a range of perspectives on the structure of low mass, \hi\ dominated galaxies, from the nearly edge-on WLM and SMC to the moderately inclined IC10, IC1613, and NGC 6822.

\begin{deluxetable*}{lcccccc}[t!]
\tabletypesize{\footnotesize}
\tablecaption{\label{tab:physical} Physical Properties of Targets}
\tablewidth{0pt}
\tablehead{
\colhead{Property} & 
\colhead{IC10} & 
\colhead{IC1613} & 
\colhead{M31} & 
\colhead{M33} & 
\colhead{NGC6822} & 
\colhead{WLM}
}
\startdata
Adopted $d$ (kpc)\tablenotemark{a} & $770 \pm 100$\tablenotemark{b} [1] & $760 \pm 36$ [2] & $776 \pm 22$ [3] & $859 \pm 24$ [3] & $526 \pm 25$ [2] & $984 \pm 19$ [4] \\
NED range in $d$ (kpc)\tablenotemark{c} & $741 \pm 85$ & $729 \pm 40$ & $780 \pm 37$ & $867 \pm 67$ & $479 \pm 31$ & $958 \pm 37$ \\
\hline
WISE1 $\log_{10} (M_\star / {\rm M_\odot})$\tablenotemark{d} & $8.7$ & $7.8$ & $10.6$ & $9.6$ & $8.1$ & $7.4$ \\
Literature $\log_{10} (M_\star / {\rm M_\odot})$\tablenotemark{d} & $7.9$ [5] & $8.0$ [5] & $10.85$ [6] & $9.5$ [5] & $8.1$ [5] & $7.7$ [5] \\
$\log_{10} (M_{\rm HI} / {\rm M_\odot})$\tablenotemark{e} & $7.88$ & $7.80$ & $\mathbf{9.80}$ & $9.30$ & $8.25$ & $7.91$ \\ 
$\log_{10} [{\rm SFR} / {\rm (M_\odot~yr^{-1})}]$\tablenotemark{f} & $-1.7$ & $-2.2$ & $-0.4$ & $-0.5$ & $-2.1$ & $-2.4$ \\
$12+\log_{10} {\rm O/H}$\tablenotemark{g} & 8.37 & 7.73 & 8.94 & 8.59 & 8.23 & 7.83\\
$\frac{d}{dR}\log_{10} {\rm O/H}$ (dex/kpc)\tablenotemark{g} & \nodata & \nodata & $-0.024$ & $-0.037$ & $-0.17$ & \nodata \\
\enddata
\tablenotetext{a}{Current best-estimated distances. References: [1] 
\citet{dellagli2018,sanna2008,gerbrandt2015}, [2] \citet{lee2024dist}, [3] \citet{Savino2022}, [4] \citet{lee2021dist}. For reference LITTLE THINGS \citep{Hunter2012} adopted distances of 0.7 kpc to IC10, 0.7 kpc to IC1613, 0.5 kpc to NGC6822, 1.0 to kpc to WLM. \citet{McConnachie2012} and \citet{Putman2021} adopted $794 \pm 44$ to IC10, $755 \pm 42$ to IC1613, $459 \pm 17$ to NGC6822, and $933 \pm 34$ to WLM.
}
\tablenotemark{b}{IC 10 has the most uncertain distance among our targets. Our adopted value is from \citet{dellagli2018} and intermediate between \citet{sanna2008} and \citet{gerbrandt2015} and the adopted \citet{Putman2021} and LITTLE THINGS values. The larger uncertainty reflects our estimate of the scatter among current quality distance estimates.}
\tablenotetext{c}{Median and standard deviation inferred from median absolute deviation for distances compiled by NED \citep{steer2017} with publication dates between 2000 and the last update of the compilation around 2018.}
\tablenotetext{d}{Stellar mass from integrating median WISE1 surface density profiles assuming our adopted distance and mass to light ratio $\Upsilon_\star = 0.5$~$L_\odot / M_\odot$. Comparisons to literature values from: [6] \citet{Yin2009} (M31) and [5] \citet{McConnachie2012} (all others). The higher WISE1 mass for IC10 agrees with 2MASS results \citet{Jarrett2003}. $\Upsilon_\star$ may be too low for M31 based on \citet{Telford2020}.}
\tablenotetext{e}{Total \hi\ mass measured, assuming optically thin emission, from the GBT \hi\ cubes with their larger spatial extents and accounting for flux calibration correction factors compared to the LGLBS VLA data. See \S\ref{sec:lglbs_hi_cd} and Appendix \ref{app:uvcombine}.}
\tablenotetext{f}{Approximate star formation rates estimated from GALEX FUV maps for IC 1613, NGC 6822, and WLM, corrected for Milky Way extinction. Estimated from WISE4 maps for IC10 and combined GALEX and WISE4 for M31 and M33 following \citet{LEROY19Z0MGS}.}
\tablenotetext{g}{Metallicities and radial gradients, rescaled for our adopted distances.  These assumed values favor $T_e$-based determinations of metallicities from optical spectroscopy of \ion{H}{2} regions. The central value of the metallicity is reported if there is a defined gradient, otherwise the value reports the average over the observations.
IC10: \citet{Cosens2024}; IC 1613: \citet{Bresolin2007}; M31: \citet{Zurita2012} M33: \citet{Rogers2022}; NGC 6822 \citet{Lee2006}; WLM: \citet{Lee2005}}
\end{deluxetable*}

\begin{deluxetable*}{lc}[t!]
\tabletypesize{\footnotesize}
\tablecaption{\label{tab:lit} Previous high resolution 21-cm and GHz radio continuum mapping}
\tablewidth{0pt}
\tablehead{
\colhead{Galaxy} & 
\colhead{References}
}
\startdata
IC10 & \makecell{\citet{Yang1993,Wilcots1998, Shostak1989,Namumba2019}; \\ \citet{Westcott2017}} \\
\hline
IC1613 & \citet{Lake1989,Lozinskaya2009,Silich2006a,Hunter2012} \\
\hline
M31 & \makecell{\citet{Brinks1984,Braun1990a,Beck1989,Braun1990b, Braun2009} \\ \citet{Koch2021}} \\
\hline
M33 & \makecell{\citet{Deul1987,Thilker2002,Tabatabaei2007a,Koch2018,Kam2017}; \\ \citet{White2019,Tabatabaei2022}} \\
\hline
NGC6822 & \citet{Dickel1985,Weldrake2003,deBlok2006,NAMUMBA17,Park2022} \\
\hline
WLM & \citet{Jackson2004, Kepley2007,Hunter2012,Ianjamasimanana2020,Yang2022} \\
\enddata
\end{deluxetable*}

\begin{deluxetable*}{lcccccc}[t!]
\tabletypesize{\footnotesize}
\tablecaption{\label{tab:obsprop} Observational Properties of Targets}
\tablewidth{0pt}
\tablehead{
\colhead{Property} & 
\colhead{IC10} & 
\colhead{IC1613} & 
\colhead{M31} & 
\colhead{M33} & 
\colhead{NGC6822} & 
\colhead{WLM}
}
\startdata
PGC & 1305 & 3844 & 2557 & 5818 & 63616 & 143 \\
R.A. (hours) & $0.3$h & $1.1$h & $0.7$h & $1.6$h & $19.7$h & $0.0$h \\
Dec. (deg.) & $+50^\circ$ & $+2.3^\circ$ & $+41.3^\circ$ & $+30.6^\circ$ & $-14.8^\circ$ & $-15.5^\circ$ \\
$v_{\rm LSR}$ (km s$^{-1}$) & -340 & -238 & -296 & -180 & -44 & -122 \\
C \& D fields & 7 & 7 & 49 & 13 & 3 & 7 \\
A \& B fields & 1 & 1 & 49 & 13 & 3 & 1 \\
\enddata
\tablecomments{PGC catalog number used to cross-reference to HyperLeda \citep{LEDA03,LEDA14}. Velocities are the ones used in our correlator setup (\S \ref{sub:correlator}) based on previous observations (Table \ref{tab:physical}) and the NASA Extragalactic Database. ``C \& D fields'' and ``A \& B fields'' refer to the number of fields in the mosaics for that target in those VLA configurations.}
\end{deluxetable*}

\begin{figure*}[!t]
\centering
\includegraphics[width=\textwidth]{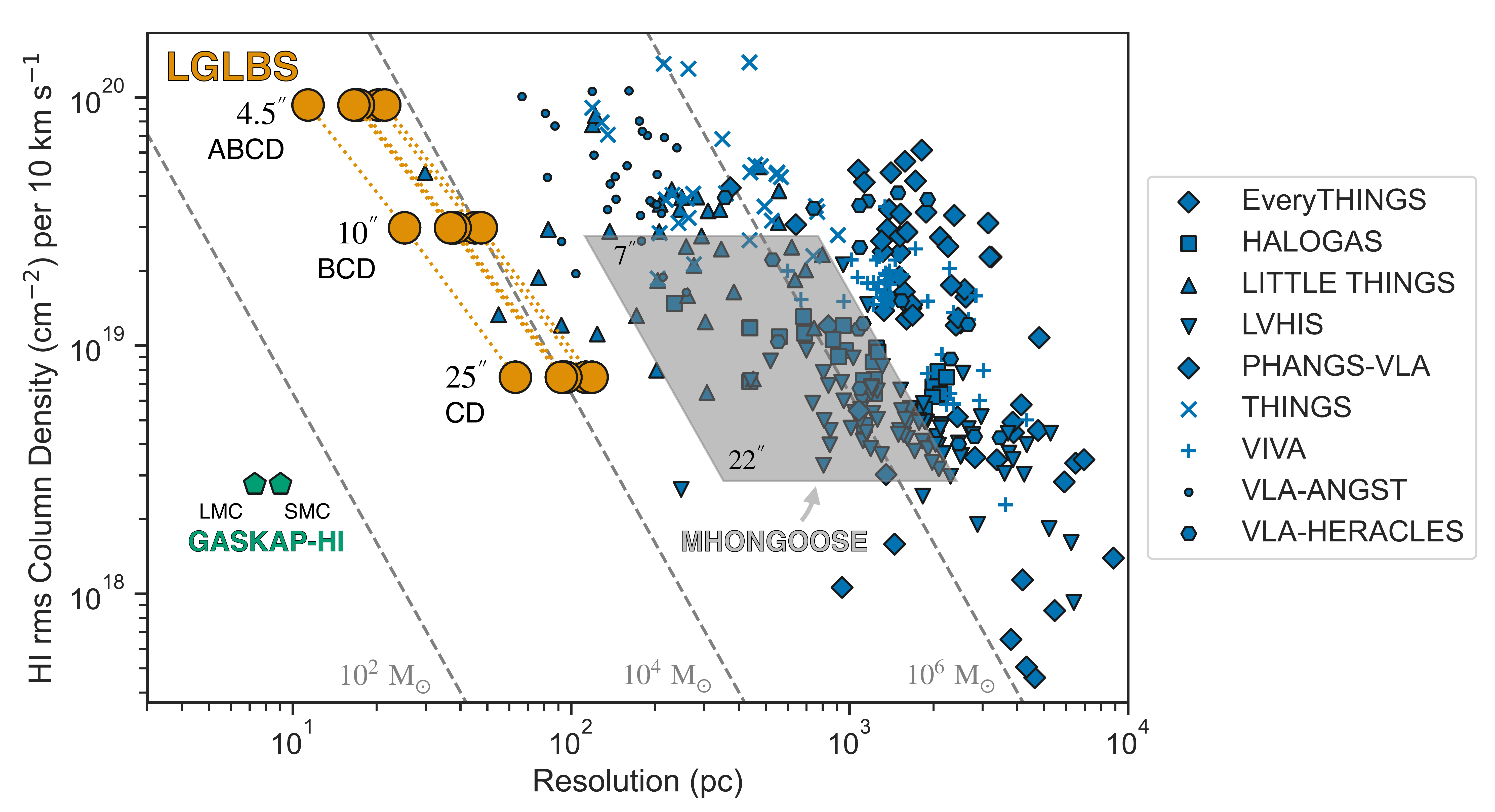}
\caption{\textbf{LGLBS \hi\ column density sensitivity and physical resolution relative to comparable nearby galaxy and Local Group surveys.} The LGLBS points show the range of sensitivity and resolution within the target velocity resolution, and the labels indicate the combination of VLA configurations combined to reach these targets.
The dashed gray lines indicate constant $5\sigma$ \hi\ mass sensitivity.
LGLBS fills an important gap in the \hi\ survey parameter space between the nearby galaxy population (e.g., 
EveryTHINGS, I. Chiang et al. in preparation;  
HALOGAS, \citealt{HALOGAS11}; 
LITTLE THINGS, \citealt{LITTLETHINGS12}; 
LVHIS, \citealt{LVHISSURVEY18};  
PHANGS-VLA, presented in \citealt{SUN22SFGAS};
THINGS, \citealt{THINGS08}; 
VIVA, \citealt{VIVASURVEY09}; 
VLA-ANGST, \citealt{Ott2012}; 
VLA-HERACLES, \citealt{Schruba2011}; 
MHONGOOSE, \citealt{DeBlok2024}), and LMC and SMC (e.g., GASKAP-HI, \citealt{Pingel2022}), which are $>10\times$ nearer than any of the LGLBS targets.
See also \citet{Maccagni2024}.}
\label{fig:lglbs_to_hi_surveys}
\end{figure*}

Because it targets the Local Group, LGLBS achieves a unique combination of sensitivity and physical resolution. Figure \ref{fig:lglbs_to_hi_surveys} shows the \hi column density sensitivity and physical resolution for LGLBS and other recent or ongoing \hi\ surveys of nearby galaxies. The proximity of our targets combined with the deep VLA coverage leads to order-of-magnitude improvements in physical resolution at comparable column density sensitivities. Thus, LGLBS bridges the gap from nearby galaxy \hi\ surveys to studies of the Milky Way and Magellanic Clouds.

Naturally, the LGLBS targets have been studied in detail in  \hi\ and radio continuum by previous surveys.  In Table \ref{tab:lit}, we list prior studies that present interferometric radio continuum and spectral line observations of our targets using the L-band. This extensive literature has informed the science case for LGLBS, though the current study supersedes these prior works in the combination of sensitivity and resolution.

\section{Observing Strategy}
\label{sec:survey}

\begin{figure*}[!t]
\centering
\includegraphics[width=0.9\textwidth]{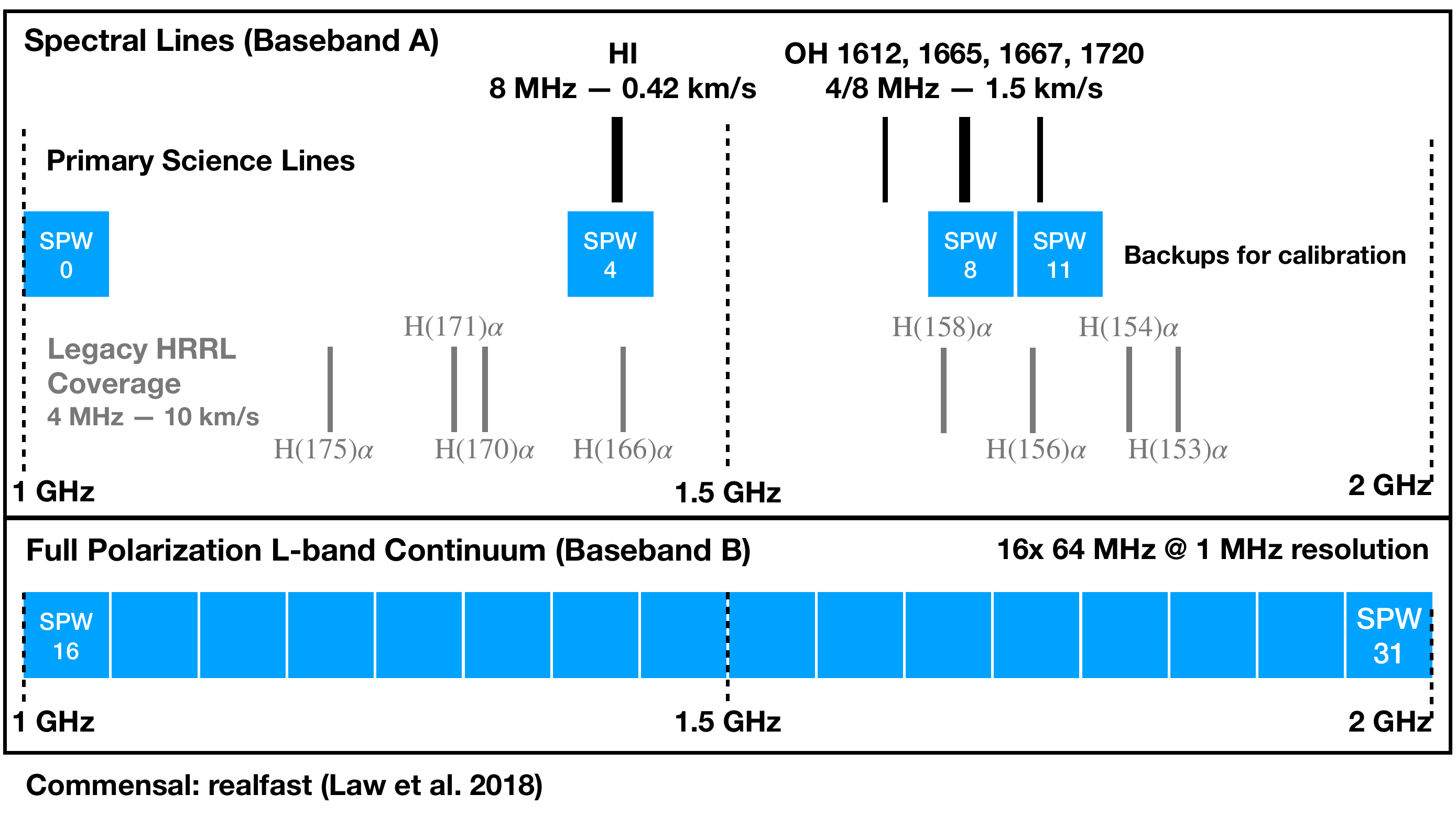}
\caption{\textbf{VLA correlator setup used for LGLBS}
(VLA Project ID 20A-346). The black, gray, and blue regions represent individual spectral windows (SPWs) used in the LGLBS setup.
The primary LGLBS data products are the \hi\ and 4 OH lines (black regions), and the Stokes I L-band continuum. Four continuum SPWs are also observed in Baseband A (labeled SPW 0, 4, 8 11 here) to allow calibration checks between the basebands. The tuning also includes full polarization for the continuum windows shown in blue (with requisite polarization calibration observations taken) and eight hydrogen radio recombination lines (RRLs, black regions). Commensal data for FRB detections was collected with REALFAST for all LGLBS observations \citep{LAW2018-realfast}. The Stokes I continuum, \hi, and OH line coverage are similar in all archival observations (Table \ref{tab:archive}) but may not exactly match the tuning used for the primary LGLBS observations. The RRL and polarization coverage vary somewhat across the archival data but are all part of all LGLBS observations.}
\label{fig:lband_tunin}
\end{figure*}

Our observations take maximum advantage of the flexible upgraded WIDAR correlator \citep[][Section \ref{sub:correlator}]{Perley2011}. We design mosaics that cover the area of known 21-cm emission and active star formation in each of our targets (Section \ref{sub:spatial-coverage}). Finally, we observe each target in each VLA configuration to provide high resolution while maintaining good surface brightness sensitivity (Section \ref{sub:sensitivity}). 

LGLBS (VLA project code 20A-346) builds on several pilot projects \citep[][]{Koch2018,Koch2020,Koch2021} that are summarized in Table \ref{tab:archive}. We integrate these data into our final data products. This pilot work adopted a similar observing strategy as presented here for the full LGLBS survey.

\subsection{Spectral Coverage}
\label{sub:correlator}

Figure \ref{fig:lband_tunin} visualizes our spectral setup. Our observations use this single L-band setup to simultaneously observe:

\begin{enumerate}

\item \textbf{21-cm \hi line:} We use an 8 MHz wide spectral band to cover the 21-cm \hi line, observing with dual polarization, 1.95~kHz channel width ($0.4$~km~s$^{-1}$), and centered at the $1420.405752$~MHz rest frequency of the 21-cm transition shifted by the recession velocity of the source given in Table \ref{tab:obsprop}. The equivalent bandwidth of 1700 km~s$^{-1}$ covers the full velocity extent of each target (\S \ref{sec:lg_sample}), as well as Milky Way emission along the line of sight, while still allowing sufficient line-free bandwidth to enable in-band continuum subtraction. The $0.4$~km~s$^{-1}$ velocity resolution ensures that we can resolve the thermal line profile of a $\sim 100$~K cloud in the cold neutral medium  ($1\sigma$ line width $\sim 0.9$~km~s$^{-1}$).
    
\item \textbf{$1-2$~GHz continuum:} We use sixteen 64 MHz spectral bands to cover the L-band continuum with 1~MHz channel width and full polarization. Our main science goals require intensity, Stokes $I$, with both polarizations averaged to minimize noise. We also require the wide bandwidth to achieve our sensitivity goals and to infer the spectral index of continuum emission. The full polarization information provides important information on the magnetic fields in our targets but represents a secondary science goal, with these products expected to follow after our initial analysis. The full polarization information is also potentially helpful for flagging radio frequency interference. We also place a few backup spectral windows into the line baseband (Baseband A). These were intended to ensure continuum coverage near key lines even in the case of dropped subbands or other correlator glitches. In the end such problems were extremely rare and these backup bands saw little use.
    
\item \textbf{OH 1612, 1665, 1667, 1720 MHz:} We observe the OH lines at 1612, 1665, 1667, and 1720 MHz. We observe the 1612 and 1720 MHz lines using 4 MHz wide spectral bands ($\Delta v \approx 720$~km~s$^{-1}$), dual polarization, 7.81 kHz ($\approx 1.7$~km~s$^{-1}$) channel width, and centered at the rest frequency of each line shifted by the recession velocity of the source given in Table \ref{tab:obsprop}. We use a single 8~MHz window with dual polarization and 7.81~kHz (1.65~km~s$^{-1}$) channel width to cover the 1665 and 1667~MHz lines together. We center this window at 1666.3804 MHz shifted by the recession velocity of the source. The width of this window means that we always cover $> 500$~km~s$^{-1}$ away from the galaxy systemic velocity. We expect to detect individual OH masers, while line stacking over large areas or absorption searches towards bright continuum sources will probe diffuse gas.
    
\item \textbf{Hydrogen radio recombination lines (RRLs):} We allocate remaining correlator resources to place bands that cover hydrogen radio recombination lines (RRLs). We cover eight transitions: H175$\alpha$, H171$\alpha$, H170$\alpha$, H166$\alpha$, H158$\alpha$, H156$\alpha$, H154$\alpha$, and H153$\alpha$. These were down-selected from the available transitions favoring regions of the spectrum free of radio-frequency interference (RFI) and avoiding potentially noisy edges of subbands. These lines are known to be faint and expected to be challenging to detect given our exposure times and the distances to our sources \citep{brown1978,beuther2016}. The data are obtained primarily as targets of opportunity and a potential archival resource for the community.

\item \textbf{Commensal REALFAST monitoring:} We worked with NRAO staff to ensure that our spectral setup was compatible with simultaneous REALFAST monitoring. REALFAST \citep{LAW2018-realfast} is a backend that processes VLA observations in parallel in real-time to monitor for transients like fast radio bursts (FRBs). Thus LGLBS also represents an extended transient search campaign towards our nearest neighbors, and \citet{Anna-Thomas2025} present 2 FRB detections towards M31 that are detected during LGLBS observations.
\end{enumerate}

The archival data listed in Table  \ref{tab:archive} includes the polarization and 21-cm line coverage that was sufficiently similar to the LGLBS strategy that there was no utility to reobserving. In many cases, the continuum spectral channels were the same in the archival data but collected in 8 128-MHz spectral windows (SPWs). While this should make little difference in principle, it did introduce some practical issues using the CASA software package to combine these data with the new 20A-346 \citep{CASA}. Details of the OH and RRL coverage varied across the different archival projects. The 21-cm coverage can all be incorporated straightforwardly into our current observations.

\subsection{Spatial Coverage}
\label{sub:spatial-coverage}

We created tailored mosaics that aim to cover the known 21-cm emission and ongoing high mass star formation in each target. These are visualized in Fig. \ref{fig:optical_collage}.

\begin{enumerate}

\item \textbf{IC10, IC1613, and WLM.} The main star-forming regions and stellar disks of these three targets fit within a single VLA primary beam (\S \ref{sec:lg_sample}), and we target one pointing in all four configurations (A, B, C, D) to achieve maximum resolution and sensitivity in this region. All three also have evidence for more extended, though often low-column density \ion{H}{1} emission \citep[][]{LITTLETHINGS12,VLAANGST12}. To capture this we observe all three targets with a 7-point hexagonal mosaic in the compact C and D configurations only. This extended coverage allows us to achieve high surface brightness sensitivity at coarser resolution over a large ($\gtrsim 10$~kpc across) area.

\item \textbf{NGC6822} has an inclined disk that is significantly extended compared to the VLA primary beam. We observe the galaxy using a three point mosaic aligned with the  \hi\ major axis.

\item We observe \textbf{M33} using a 13-pointing mosaic that covers the high column density \ion{H}{1} known from previous mapping \citep[][]{Putman2009,Gratier2010a}. This mosaic is identical to that used in \citet{Koch2018} to allow easy combination of our new observations with these previous data (Table \ref{tab:archive}).

\item We observe \textbf{M31} using a 49-pointing mosaic. This covers the high column density \ion{H}{1} known from previous mapping \citep[][]{Braun2009}, but does not cover the outer disk of the galaxy \citep[i.e., the extended low column material beyond the optical radius of M31][]{Braun2004}. More than any other target, M31's large extent forces us to focus specifically on the disk of the galaxy.

\end{enumerate}

We cover the observed area with a hexagonal grid of pointing centers, spaced by 1000\arcsec\ (1048\arcsec\ for M31), which is approximately one half of the full width at half maximum (FWHM) primary beam size at $1.4$~GHz. This coverage yields uniform rms noise across the mosaic, particularly in the lower half of the frequency band. The good overlap in coverage between adjacent pointings helps ensure even $uv$ coverage so that losing a few scans to RFI has minimal impact. In principle, covering bright continuum sources in different parts of the VLA primary beam should improve our bright source reconstruction and deconvolution.

Our scan time, calibrator cycling time, and the length of individual integrations followed standard VLA recommendations for each configuration. Each mosaic was completely observed during every observation. Given typical observation durations of 3-5~h, most mosaics were observed several times. However, each field in M31 could be observed only once per observing block. To ensure even $uv$ coverage across M31, we scheduled observations with a variety of LST start times, and we also rotated the order in which we cycled through the mosaic pointings. In \S\ref{sec:lglbs_hi_cd}, we show derived \hi\ products using the C and D configurations that demonstrate that the mosaics achieve a uniform sensitivity.

\subsection{Target Depth and Balance Between Configurations}
\label{sub:sensitivity}

As described in \S \ref{sec:science_drivers}, we aim to pursue multiple science goals with a single set of observations. These science goals drive our sensitivity and resolution requirements to achieve four main observational goals:

\begin{enumerate}

\item Measure the line-integrated column density of the \ion{H}{1} at $< 20$~pc ($\approx 5''$) scales and reach rms column density sensitivity of $\approx 1 \times 10^{20}$~cm$^{-2}$. This is sufficient to measure the column density of individual clouds and identify \ion{H}{1} shells and holes.

\item Measure high surface brightness, high velocity resolution spectra of the 21-cm line at $\approx 40{-}60$~pc ($\approx 10{-}15''$) scales to allow detailed spectral analysis. We target $\approx 4$~K per $0.4$~km~s$^{-1}$ channel.

\item Detect and resolve continuum emission from individual supernova remnants and \ion{H}{2} regions, which have sizes of $\sim 10$~pc ($\approx 2.5''$) and fluxes of $\gtrsim 10\mu$Jy.

\item Recover the full 21-cm emission at all spatial scales accessible to the VLA, giving a complete view of the atomic gas in each galaxy.

\end{enumerate}

Our observing strategy balances these goals and targets effective per-point integration times by configuration of 11 h in A, 11 h in B, 11 h in C, 5.5 h in D. We account for the oversampling of the mosaics when working out the target time per field. The high resolution goals, especially the continuum, push us to weight the high resolution VLA A configuration heavily. Those emphasizing surface brightness sensitivity, especially in narrow velocity channels, benefit most from the intermediate B and C configurations. Sensitivity to extended emission requires the D configuration.

Our adopted balance of time between different configurations differs somewhat from the ``rule of thumb'' that prescribes a 3:1 ratio between successive configurations and was used by previous \ion{H}{1} surveys \citep{THINGS08,LITTLETHINGS12,VLAANGST12}. This reflects that our A configuration science drivers are heavily weighted towards continuum, which has less stringent surface brightness sensitivity requirements than 21-cm imaging and benefits from larger bandwidth. It also reflects a practical acknowledgment that fully leveraging the high spectral resolution (which divides the line signal into many dozens of individual channels) requires emphasizing the intermediate configurations. We validated that this balance of configurations yields reasonable PSFs using simulations, and our resulting imaging closely matches the simulated expectations.

\subsection{Short and Zero Spacing Data}
\label{sub:shortspacing}

Our targets all show significant extent compared to the largest angular scales to which the VLA C and D configurations are sensitive. This largest recovered scale is $20.5\arcmin$ for the 21-cm \hi line, and ranges from $29.4$\arcmin\ to $14.7$\arcmin\ from 1 to 2 GHz across the L-band continuum. For the 21-cm line, GBT observations exist for all of our targets at an appropriate depth to serve as short- and zero-spacing observations. The individual projects are listed in Table \ref{tab:archive}, and thanks to the upgraded GBT VEGAS backend, most have a velocity resolution appropriate for combination with our VLA observations. We describe the reprocessing and imaging of the GBT observations for NGC6822, IC10, IC1613, and WLM in Appendix \ref{app:gbt_data}.  In Appendix \ref{app:uvcombine}, we describe the short-spacing correction by feathering and relative flux comparison tests for the GBT and VLA 21-cm \hi\ observations.

In principle, the continuum observations also require short- and zero-spacing corrections. This problem is more acute than for the spectral lines since the Doppler shift across the galaxies means the line emission only appears in a spatially confined region while continuum emission spans the entire galaxy. Unfortunately, stable, high quality L-band single dish continuum mapping is challenging \citep[e.g.,][]{Beck1998} and has been almost exclusively restricted to narrower frequency ranges to avoid instability from RFI. Identifying or obtaining a suitable set of zero spacing maps that span the full 1{--}2 GHz range remains a future goal. For now, our continuum science emphasizes compact scales and individual objects. Individual \ion{H}{2} regions or supernova remnants have sizes far smaller than the largest angular scales recovered by the C and D configuration and should be robustly imaged by using our VLA data alone.

\begin{deluxetable}{lccc}[t!]
\tabletypesize{\footnotesize}
\tablecaption{Data sets included in LGLBS\label{tab:archive}}
\tablewidth{0pt}
\tablehead{
\colhead{Galaxy} & 
\colhead{Config} & 
\colhead{Program ID} & 
\colhead{PI}
}
\startdata
All & A/B/C/D & 20A-346 & Leroy \\
\multicolumn{4}{c}{co-PIs: Chomiuk, Dalcanton, Rosolowsky, Stanimirovic, Walter} \\
\hline
IC1613 & C & 13A-213 & Leroy \\
\hline
M31 & D & 14A-235 & Leroy\\
M31 & C & 15A-175 & Leroy\\
M31 & B & 15A-175 & Leroy \\
\hline
M33 & C & 14B-088 & Rosolowsky \\
M33 & B & 17B-162 & Koch \\
M33 & A & 16B-232 & Koch\\
M33 & A & 16B-242 & Koch\\
\hline 
NGC6822 & DnC & 13A-213 & Leroy\\
NGC6822 & DnC & 14B-212 & Schruba\\
NGC6822 & CnB & 14B-212 & Schruba\\
NGC6822 & CnB$\rightarrow$B & 14B-212 & Schruba\\
\hline
WLM & DnC & 13A-213 & Leroy\\
\hline
IC10 & GBT & 13A-430 \& 13A-420 & Ashley \\
IC1613 & GBT & 16A-413 & Pisano \\
M31 & GBT & 14A-367 & Leroy \\
M33 & GBT & 09A-017\tablenotemark{a} & Lockman \\
NGC6822 & GBT & 13B-169  & Johnson \\
WLM & GBT & 16A-413 & Pisano \\
\hline
\enddata
\tablecomments{20A-346 represents the main survey. The other projects are archival projects which target the sample galaxies with similar spectral setups since the VLA upgrade.} 
\tablenotetext{a}{\citet{LOCKMAN2012}.}
\end{deluxetable}

\section{Data Reduction \& Quality Assurance}
\label{sec:dr_qa}

Although the VLA calibration pipeline works well, the presence of substantial RFI in L-band and occasional issues with antenna performance mean that most tracks require additional flagging and quality assurance (QA) checks before imaging. To deal with the substantial data volume of LGLBS, the need for rapid review of observations, and the distributed nature of our team, we developed a custom QA system.
This system is publicly available and comprised of the \texttt{ReductionPipeline}\footnote{\url{github.com/LocalGroup-VLALegacy/ReductionPipeline}} and \texttt{QAPlotter}\footnote{\url{github.com/LocalGroup-VLALegacy/QAPlotter}} packages. These packages can easily adapted to process any non-LGLBS VLA L-band data taken using the WIDAR correlator.

Our system expands upon the quality assurance and flagging capabilities in the current VLA pipeline, adding additional interactive capabilities that can be used to identify poor visibility data. Our system stores all data in a single compute facility, and team members perform QA review remotely using a web browser. Thus, a team member never needs direct access to the visibility data. Users specify flagging commands to eliminate bad data from subsequent pipeline calibration. The system then implements those manual flags, reruns the calibration, then serves up new interactive plots, allowing users to iterate until all bad data are removed from the calibration and the resulting calibration appears robust. 

\subsection{Data Reduction}
\label{sub:reduction}

Our procedures build on the VLA pipeline\footnote{\url{https://science.nrao.edu/facilities/vla/data-processing/pipeline}}. As observations were taken over multiple years, we updated to use new pipeline versions as they were released. Our calibration used CASA VLA Pipeline versions ranging from 6.1.2-7 to 6.5.4-9.

As recommended in the NRAO documentation, we split the data into separate continuum and spectral line measurement sets prior to running the reduction pipeline. This allows us to apply Hanning smoothing to the continuum data, which minimizes Gibbs ringing due to strong RFI. We do not smooth the spectral line data, which are instead processed at their native resolution. Narrow \hi\ features on scales similar to the $\sim0.4$~\kms channel width are important to our science goals (\S\ref{sec:science_drivers}) and these would be degraded by smoothing. Fortunately, RFI is typically weaker in our \hi\ data because our source redshifts fall within the protected part of the spectrum.

Most of the pipeline steps calibrate per spectral window (SPW), so splitting the data has minimal effect on the calibration solutions. The exception is flux bootstrapping between calibrators. For each calibrator, the VLA pipeline calculates the average flux in each SPW and then fits these together to constrain the spectral index and flux. Then, the pipeline bootstraps from the flux calibration standard to infer the flux of the secondary calibrators (i.e., those used to measure gain, polarization angle, and leakage) based on the results of these fits. Because the spectral coverage varies between the continuum and spectral line parts, the bootstrapped fits could, in principle, be different. However, based on reviewing a subset of tracks, we find differences of $<5\%$ between the boostrapped fluxes for the same calibrator in the line and continuum parts.  This is within the expected absolute flux uncertainty for the VLA at L-band\footnote{\url{https://science.nrao.edu/facilities/vla/docs/manuals/oss/performance/fdscale}}.

\subsubsection{Continuum Data Reduction}
\label{subsub:continuum_pipeline}

The VLA pipeline is optimized for continuum calibration, and we follow the standard pipeline recipe for continuum data. For L-band, the heuristics in the pipeline work well for flagging strong RFI and determining where calibration solutions have failed. However, we still find it useful to add two flagging routines to the standard pipeline. First, we flag the 1 or 2 continuum channels closest to the target's \hi\ emission in order to remove potential bias due to line contamination in the continuum images. We found that when the VLA is in the compact C and D configurations, the \hi\ emission was strongly detected in the continuum channels for many fields in M31 and M33. Second, we moderately increase the time interval in the \texttt{quack} RFI flagging routine to flag the first 3 integrations of each scan. This flagged many cases where a subset of antennas began observations before others, which appears as integrations with sparse $uv$ coverage at the start of the science scans after applying the observatory-generated flags.

\subsubsection{Spectral Line Data Reduction}
\label{subsub:line_pipeline}

We make several modifications to the standard VLA pipeline to optimize it for the LGLBS spectral line setup. First, following the VLA pipeline guide, we automatically specify line frequency ranges where we expect astronomical signal. This helps avoid the case where the automated RFI flagging routines confuse spectral-line features with RFI and flag them by accident. The specific protected spectral ranges are defined per galaxy based on the velocity ranges in the kinematic LSR (LSRK) frame known to exhibit \hi emission. We also include a protected range near 0~\kms in all of our target fields to avoid flagging foreground Galactic \hi emission and absorption.

For the spectral line data on the science targets we also disable some automated RFI flagging routines altogether. We found that the built-in heuristics, which have been primarily tested on continuum SPWs, flagged very few data in comparison to the amount of time needed to run these routines. We note that this might not be an optimal choice for other line observations. Our Local Group targets have small systemic velocities, so the source emission still falls within protected radio-quiet frequency ranges. These small Doppler shifts result in minimal RFI in the \hi\ and OH line (1665/1667 MHz) windows.

We also add capabilities to handle the Galactic \hi foreground towards the calibrators, where the intervening gas creates absorption features in the calibrators' continuum spectra. For each calibrator, we define and flag a velocity range based on \hi absorption spectra where available \citep[e.g., following 21-SPONGE;][]{Murray2015}. When a \hi\ absorption spectrum is not available, we examine the \hi spectral window from the first observations and define a velocity range based on the observed \hi absorption. Avoiding Galactic \hi absorption is crucial for bandpass calibration and further directly impacts \hi measurements in our science targets, as M31 and NGC6822 overlap in velocity with Galactic \hi. To recover this velocity region for the bandpass solution, we interpolate across the flagged gap using a Savitzky-Golay filter with handling for complex values. This approach is stable because, for the 8~MHz bandwidth that we use for the \hi , the bandpass response of the VLA varies smoothly on scales much larger than the Galactic \hi\ velocity width. This procedure is not needed for the OH lines, which do not show obvious Galactic absorption features at a level that affects our calibration procedures.

\subsection{Quality Assurance}
\label{sub:qa}

We found that despite the good performance of the pipeline, residual RFI was often present and other technical issues also persisted after the automated calibration. In practice, we found that the best practical way to address this was manual inspection of the calibrated data set produced for each 3-5 long scheduling block. During this manual inspection, the reviewer identifies egregious or systematic issues affecting the calibrated data or the calibration and then creates manual flagging commands to remove these issues. After inspection, subsequent pipeline runs use these manual flagging commands. Team members then inspect the newly calibrated data using these additional flags, repeating the process until we have final, well-calibrated visibility data.

\subsubsection{Interactive QA Plots}
\label{subsub:interactive_qa_plots}

Our distributed QA system uses a web browser to provide a visualization interface. This includes bespoke interactive visibility and calibration plots and an automated review and restart procedure built around a series of Google Sheets linked to CASA scripts. Specifically, we load the data into the CASA task \texttt{plotms} and produce a series of projections of the visibility data that are exported as text files. These summary files are automatically created as the last step in each pipeline run\footnote{We built standard export functions for each interactive plot type that can be used with most VLA data sets that have been reduced with recent versions of the VLA reduction pipeline. These routines assume naming conventions adopted for the VLA pipelines associated with CASA versions $>6.1$.}.
After the pipeline run, a separate package (\texttt{QAPlotter}\footnote{Similar to the CASA export step, much of the \texttt{QAPlotter} implementation can be adapted for use with other interferometric data from different telescopes.}) imports these data and collates the quick look images used for QA into interactive figures using \texttt{astropy.table} and \texttt{plot.ly}. The HTML figures created by \texttt{plot.ly} have interactivity including panning and zooming.
Finally, the \texttt{QAPlotter} software exports the HTML plots to a webserver that is available via a browser to any team member and embeds them in an HTML framework that makes it easy to navigate between the different QA plots.

\begin{figure*}[!t]
\centering
\includegraphics[width=\textwidth]{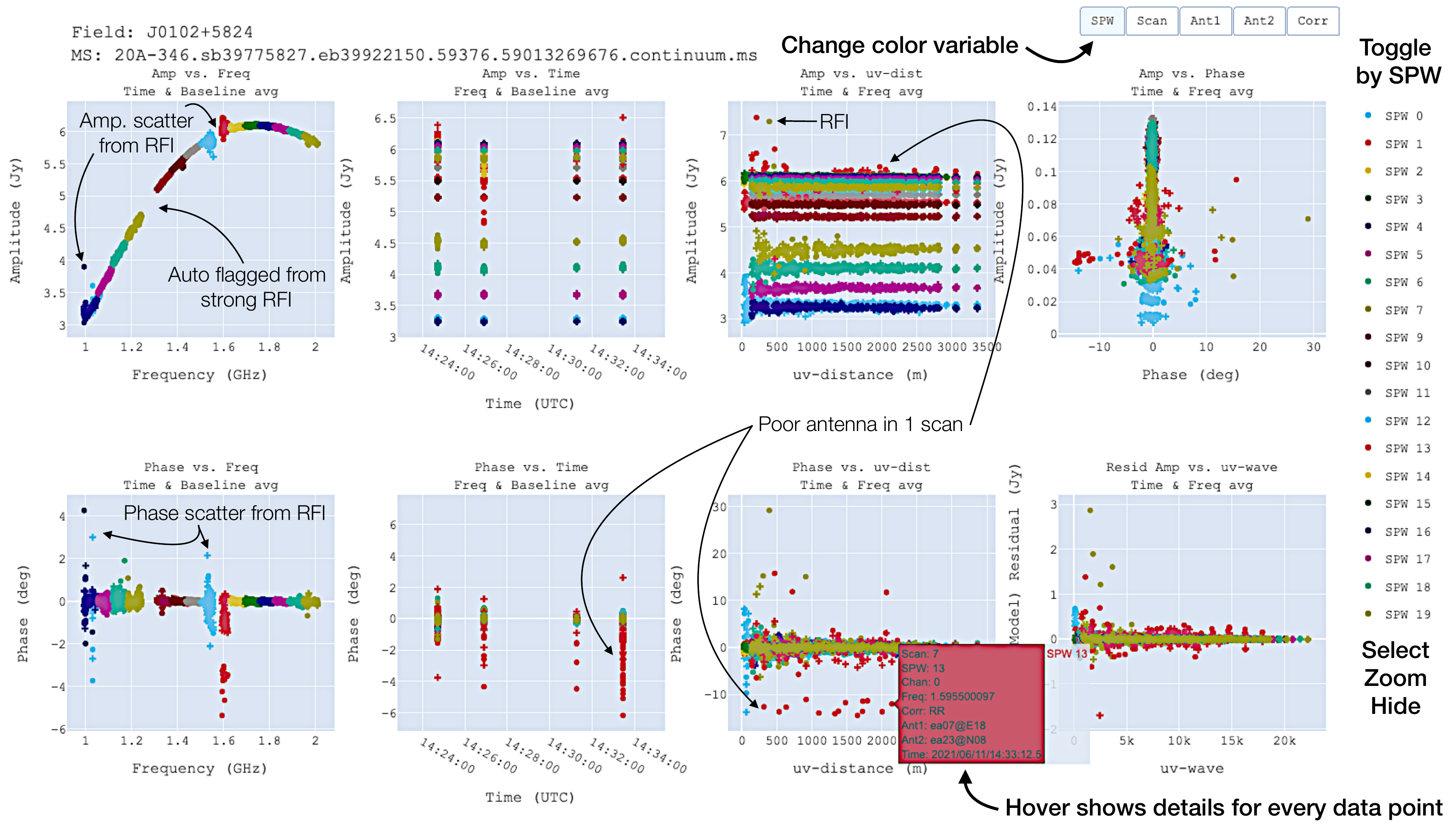}
\caption{\textbf{Example of an interactive plot used to quality assure LGLBS visibility data.} Calibrated visibilities for the gain calibrator J0102+5824 (used for NGC6822 observations), with plots showing average amplitude and phase as a function of time, frequency, antenna, and $uv$ distance.
This plot shows the initial calibrated visibilities from the pipeline without additional manual flagging applied.
We indicate several cases of additional RFI that was not automatically flagged, and an example of a set of poor amplitude/phase baselines arising from a single antenna in 1 scan.
In bold text, we highlight features of the interactive plots that enable rapid identification of visibility errors, including the interactive key where data from individual SPWs can be toggled to select, hide, or zoom to a subset of the data.
}
    \label{fig:qa_plots}
\end{figure*}

\begin{figure}[!t]
\centering
\includegraphics[width=0.5\textwidth]{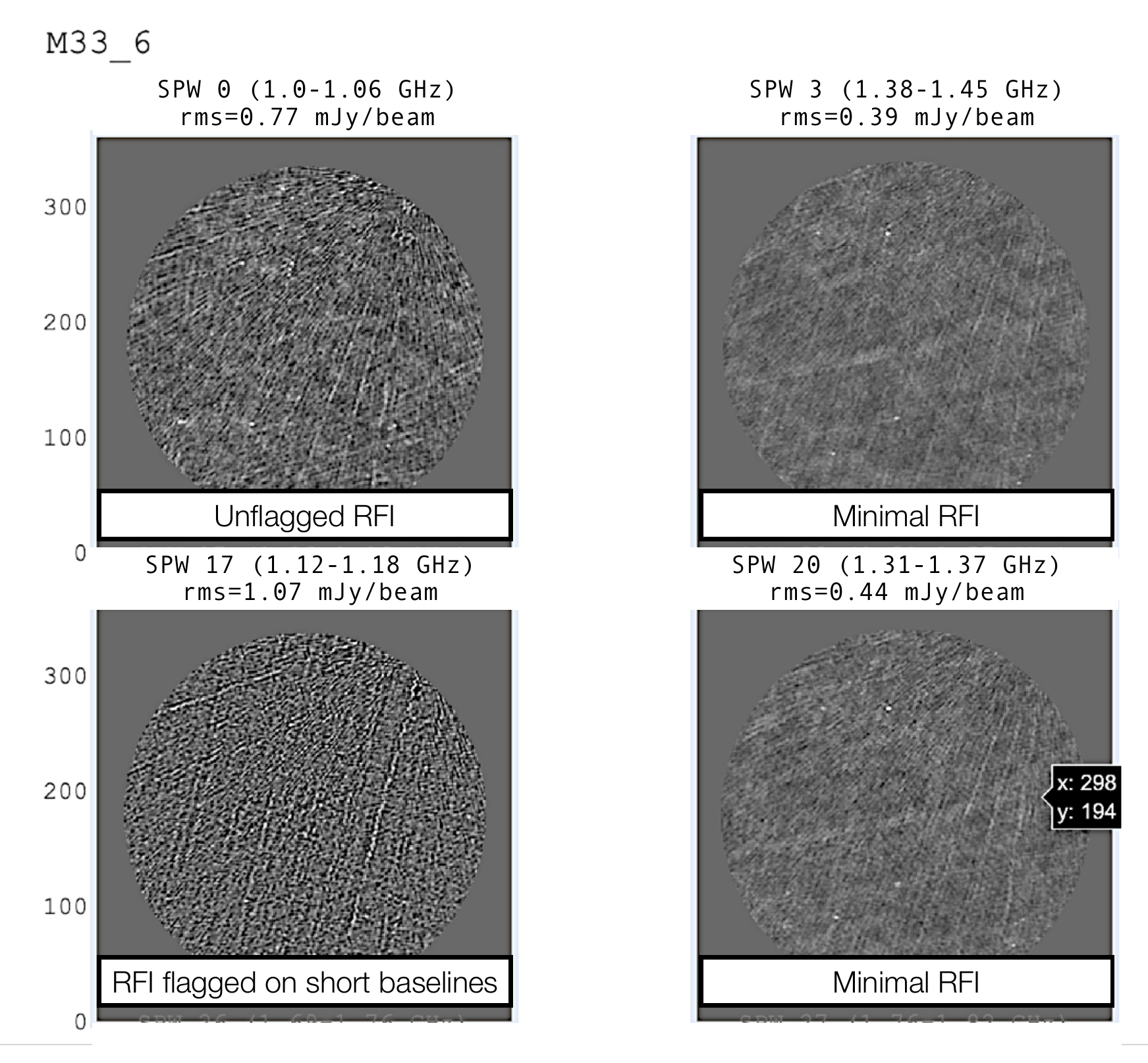}
\caption{\textbf{A subset of quicklook imaging for a single M33 field examples used to quality assure LGLBS visibility data}, which the LGLBS pipeline generates for every science field.
We show 4 examples generated from single 64~MHz continuum SPWs with no deconvolution applied.
The left column shows examples of poor data quality, due to either remaining RFI that requires flagging (top) or where strong RFI has been automatically flagged by the VLA pipeline (bottom).
The right column are examples of SPWs with minimal strong RFI remaining, where the artifacts are consistent with the VLA's point-spread-function shape.
}
    \label{fig:qa_plots_quicklook}
\end{figure}

Our goal in manual QA is for a team member to identify and flag calibration or data issues that had been missed by existing algorithmic or heuristic-based measures. To enable this we create three types of interactive plots:

\begin{enumerate}

\item Summary plots showing the final calibration tables, including the amplitude and phase bandpass tables, and the phase and amplitude gains as a function of time.
    
\item Visibility summary plots per calibrator and science field, averaging over combinations of frequency, time, and $uv$-distance to minimize data volume. For science fields, we create panels showing amplitude versus time, frequency, and $uv$-distance. For calibrators, we also include panels showing phase as a function of time and frequency, as well as amplitude and phase versus antenna number. The latter allows us to identify systematic issues per antenna. Figure \ref{fig:qa_plots} shows an example visibility summary plot for the gain calibrator J0102+5824.

\item Quick look imaging of each SPW for each field with no deconvolution (also known as the ``dirty image''). This allows us to quickly assess the imaging quality. It also allows us to (roughly) assess the achieved noise level and compare it to the theoretically expected noise. We analyze these quick look images to create interactive scatter plots of the noise per SPW and per field. Anomalously high noise levels proved useful to quickly identify poorly calibrated visibilities. Figure \ref{fig:qa_plots_quicklook} shows an example of quicklook continuum images for a single field in M33.
    
\end{enumerate}

Each of these interactive plots retain sufficient metadata to identify the frequency, time, antenna, baseline, and polarization associated with outlier data or other issues. These metadata also enable changing the color of points in the visibility summary plots to expose different data axes (e.g., users could quickly switch between coloring by polarization or SPW in an amplitude vs.\ time plot). Finally, we also include the ability to focus on individual chunks of the data through interactive legends; for example, selecting only a subset of SPWs or fields to be shown in a visibility summary plot is possible.

\subsubsection{Common Issues Identified in QA Process}
\label{subsub:qa_procedure}

Using these tools, we have quality assured more than $430$ individual tracks ($\sim1800$~h; including archival VLA observations) as of this writing, distributing the work among $21$ team members\footnote{QA participants include: Haylee Archer, Alberto Bolatto, Katie Bowen, Michael Busch, Hongxing Chen, Serena Cronin, Harrisen Corbould, Jeremy Darling, Thomas Do, Jennifer Donovan Meyer, Cosima Eibensteiner, Marcus Graham, Deidre Hunter, Eric Koch, Timea Kovacs, Lauren Laufman, Adam Leroy, Jarod McCoy, Amit Kumar Mittal, Hailey Moore, Nickolas Pingel, Erik Rosolowsky,  Daniel Rybarczyk, Sumit Sarbadhicary, Amy Sardone, Snezana Stanimirovic, Ioana Stelea, Jiayi Sun, Elizabeth Tarantino, and Vicente Villanueva.} and iterating each track until it is properly calibrated. Another $60$ tracks, all targeting M31 in A configuration, are currently undergoing iteration in QA.

Since QA effort is distributed over a broad group, we began the QA process with a smaller team with significant previous experience with interferometry data reduction to produce a ``best-practices'' QA guide to help ensure that our review process was uniform among reviewers. Based on the aggregate statistics shown below in \S\ref{subsub:qa_flagstats}, we do not find significant differences in the flagging statistics after QA between individual reviewers.

While the vast majority of systematic issues are correctly caught and handled with the VLA pipeline heuristics, we highlight some common issues in the LGLBS data identified during this experience. We expect that continued development of the pipeline will further reduce the need for manual flagging.

At L-band, particularly for continuum data, the most common issue flagged is transient RFI at persistent frequencies. In practice, our reviewers mostly manually flagged RFI that appeared as $>0.1$~Jy in amplitude vs. frequency plots created after averaging over the duration of a scan ($\sim3-5$~min) and all antennas. We verified that this flagging does moderately lower the estimated noise in the resulting quick look images. 

The QA process also frequently identified poorly performing antennas that show anomalously low amplitudes or individual outlying baselines. Poorly performing antennas were frequently evident in the bandpass table, where amplitudes $>3\times$ lower than the median of the remaining antennas leads to visible increase in the scatter in subsequent gain calibration solutions. Many of these per-antenna or per-baseline issues correspond to issues noted by operators in the observing logs but not flagged automatically by the pipeline. We also identified some rarer issues, including a handful of cases where single antennas have anomalously low amplitudes ($>100\times$ lower but nonzero), but only in individual scans.

Finally, we note that for the OH 1612 MHz line, significant velocity ranges are often lost due to persistent, strong RFI near 1615~MHz. In our survey, this most significantly affects M31, where the blue-shifted Southern half of the galaxy with recession velocity from $v_{\rm LSRK} \sim-300$ to $-500$~\kms falls in this frequency range. In these cases, we remove the most egregious RFI frequency ranges and attempt to recover as much of the red-shifted half of M31 as the data allow.

The manual flagging produced through our QA procedure offers a valuable set of labels for training machine-learning approaches to identify common issues in interferometric data. The data have uniform spectral coverage, common processing, and repeated visits to every field over $\sim1$ to 10 year time scales (including the archival projects; see Table \ref{tab:archive}), making them an ideal reference set.  The next generation of radio interferometers \citep[e.g., the SKA or ngVLA,][]{SKA, Murphy2018} will require extending these manual or automated procedures to cope with the expected data rates.

\subsubsection{Flagging Statistics after QA}
\label{subsub:qa_flagstats}

Here we quantify the overall flagging fractions based on a representative sample of 337 LGLBS tracks (all with project code 20A-346) reflecting the qualitative description of common QA issues.

\begin{figure}[!t]
    \centering
    \includegraphics[width=0.48\textwidth]{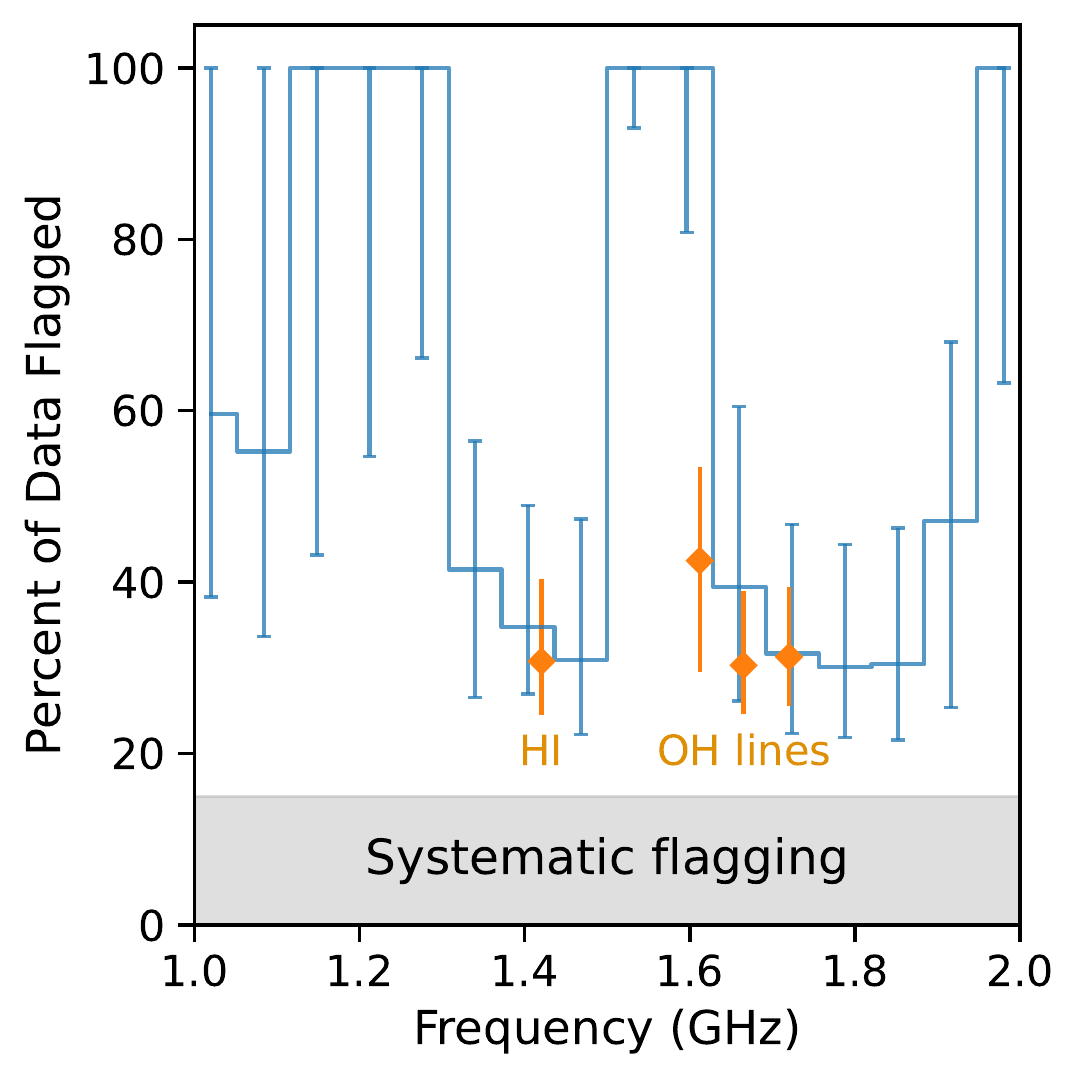}
    \caption{\textbf{Median percent of flagged data per SPW, where the errorbars reflect the 15$^{\rm th}$ to 85$^{\rm th}$ percentile range of 337 LGLBS tracks}. The continuous blue lines indicate flagging percentages for the continuum SPWs, and the orange diamonds are the line SPWs for \hi and the 4 OH lines (the OH1665/1667 are recorded together in the same SPW). The current LGLBS processing does not include the Hydrogen RRLs included in the spectral setup.}
    \label{fig:flagging_by_freq}
\end{figure}

\begin{figure*}[!t]
    \centering
    \includegraphics[width=0.9\textwidth]{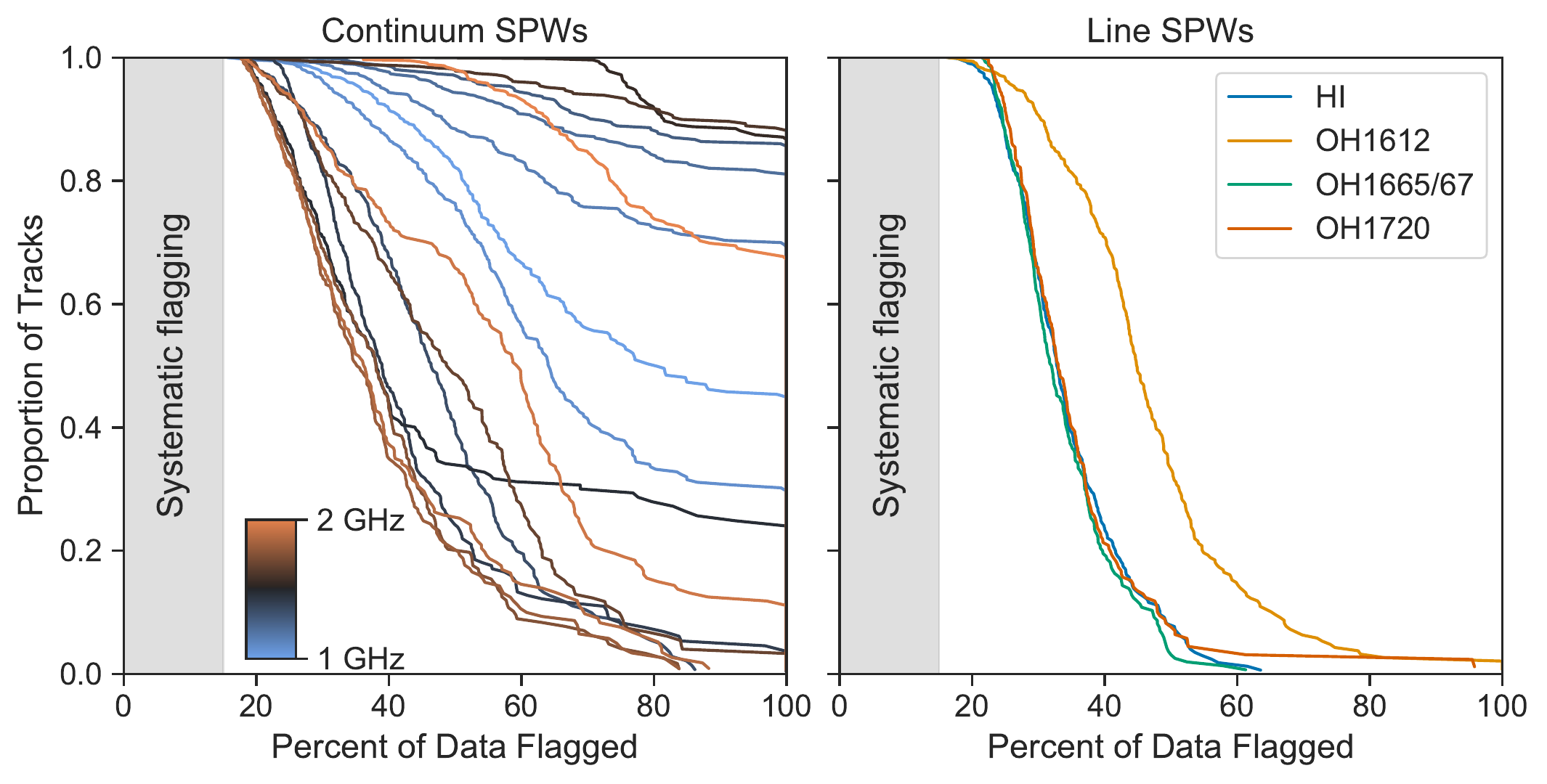}
    \caption{\textbf{Cumulative distribution functions (CDFs) of the  percent of flagged data split by SPW over 337 LGLBS tracks.} Left: Flagging percentage CDFs for the 16 64~MHz continuum SPWs covering 1--2~GHz. Right: Flagging percentage CDFs for the \hi and 4 OH line SPWs.}
    \label{fig:flagging_cdfs}
\end{figure*}

Figures \ref{fig:flagging_by_freq} and \ref{fig:flagging_cdfs} summarize the flagging percentages across the continuum and spectral line SPWs for the LGLBS targets using 337 LGLBS tracks that provide representative statistics for the whole survey.
As indicated on both figures, $\sim15\%$ of the flags result from systematic flagging (channel edges, first 3 integrations, etc).
We calculate the flagging percentages in each track using the statistics saved in the VLA pipeline weblog, which we average across antennas and fields in each track to measure the total flagged percentage in each SPW.
Figure \ref{fig:flagging_by_freq} shows the median flagging percentage versus frequency, where the errorbars indicate the 15-to-85$^{\rm th}$ percentiles across the 337 tracks.

Consistent with RFI spectra at the VLA site, Figure \ref{fig:flagging_by_freq} shows that we typically fully flag continuum SPWs from 1.1 to 1.3~GHz and 1.5 to 1.6~GHz, with the lowest data losses between 1.3-1.45 and 1.6-1.9~GHz.
The spectral line SPWs have similar or moderately lower flagging fractions relative to the continuum SPWs, consistent with the location of these lines in protected bands near the rest frame (for \hi, OH1665, OH1667) or in minimal-RFI contaminated ranges (OH1720).
As noted above, the OH1612 rest frequency falls near to strong and persistent RFI, though only the redshifted edge of OH1612 for M31 significantly overlaps with this strong RFI range.

Figure \ref{fig:flagging_cdfs} shows the cumulative distribution functions (CDFs) of data flagging percentages for the continuum and line spectral windows. In general, these highlight the trends visualized in Figure \ref{fig:flagging_by_freq} but also show the full distributions of SPW flags in the sample of 337 tracks. For the continuum SPWs, the CDFs show the high loss fraction for SPWs with strong and persistent RFI (particularly near 1.5~GHz). For the line SPWs, we find comparable statistics for \hi and OH1665/67, consistent with much of our bands falling within the protected frequency ranges. OH1720 has similar flagging statistics, reflecting the minimal RFI at these frequencies, but OH1612 has a consistently higher flagging fraction because of its proximity to strong RFI sources. For each of the lines, we note that the aggregate flagging statistics reflect the whole SPW bandwidth, while our target velocity ranges typically fill $20\mbox{--}50\%$ of the SPW.  Hence, the line flagging statistics present a pessimistic view of the spectral line sensitivity as RFI preferentially occurs closer to the edge of the protected bands (see \S\ref{sub:ic10_noise_mass_sensitivity}).

\subsection{Data Products}
\label{subsub:data_products}

In later LGLBS data releases, we will deliver the final calibration tables and flag versions of each track to the NRAO archive, allowing other users to regenerate calibrated visibility data. A final description of these products will accompany each LGLBS data release.

For our own processing, after a track passes QA, we split the calibrated visibilities of the science targets from the original measurement sets to minimize the data volume. The calibrated science data are then transferred to persistent storage to be accessed for subsequent imaging using separate approaches for spectral lines (Pingel et al.\ in prep.) and continuum (Sarbadhicary et al. in prep). 

\subsection{\hi\ resolution and sensitivity achieved in LGLBS}
\label{sub:ic10_noise_mass_sensitivity}

\begin{figure*}[!t]
    \centering
    \includegraphics[width=1.0\textwidth]{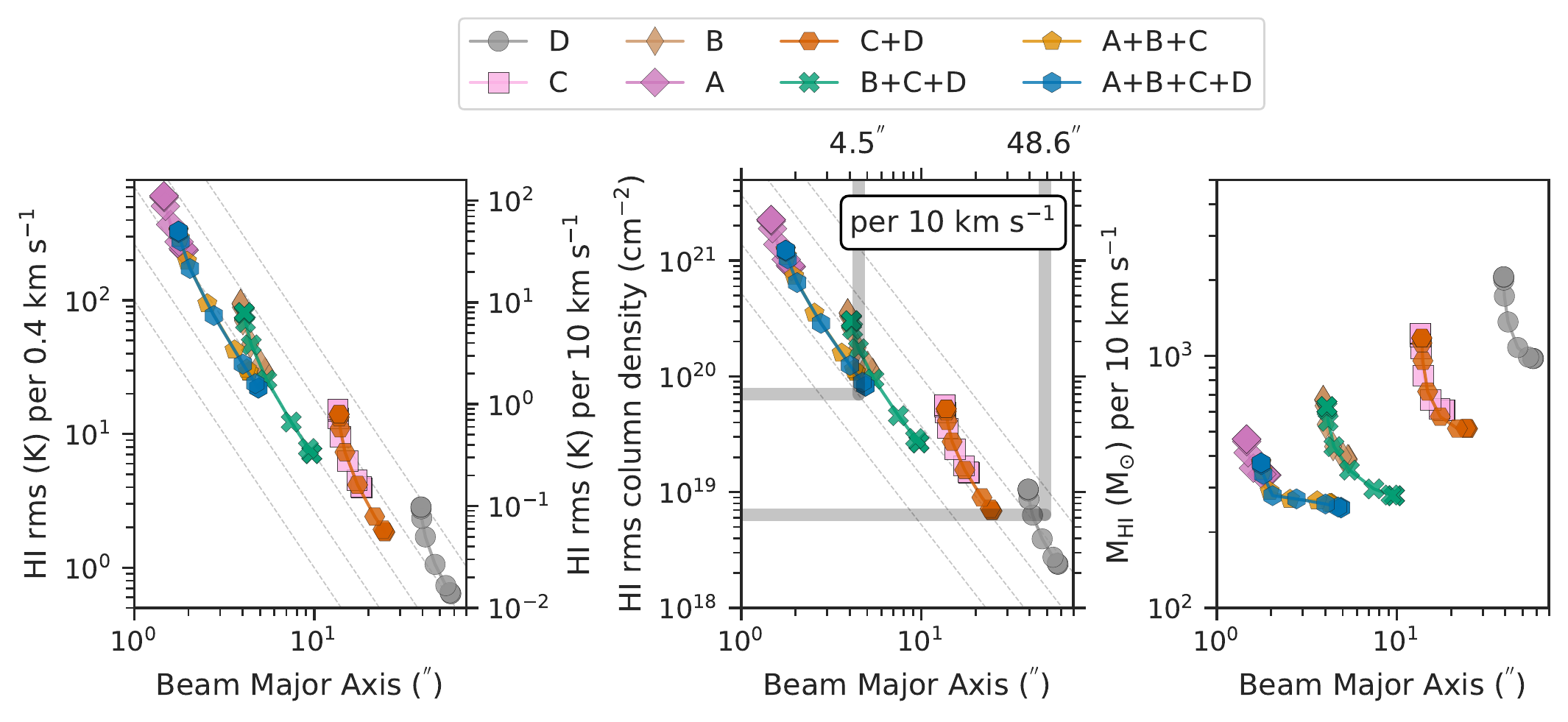}
    \caption{\textbf{Measured \hi\ intensity (left), column density (center), and mass per beam sensitivity (right) using observations of our central IC10 pointing.} Each color corresponds to a different array configuration or combination of array configurations (see legend). The points connected by lines for each color show the effect of varying the weighting scheme used to grid the data (from uniform, through robust with Briggs parameter $-2$ to $2$, to natural weighting). The left panel shows the rms noise for 21-cm intensity in an individual $\Delta v = 0.4$~km~s$^{-1}$ channel. The middle panel shows the implied column density sensitivity given the measured noise and integrating over a 10~km~s$^{-1}$ width. The right panel shows the mass sensitivity per beam, also assuming a 10~km~s$^{-1}$ velocity integration window. The gray dashed lines in the left and center panels indicate the expectation from statistical averaging of a constant noise map. Deviations from this reflect the loss of sensitivity when fewer baselines are included at coarse resolution. The solid gray lines in the center panel  note the target resolution and sensitivity goals for the survey at the finest (4.5\arcsec at $7\times10^{19}$~cm$^{-2}$) and coarsest (48.6\arcsec at $6\times10^{18}$~cm$^{-2}$) scales. We reach within $\sim20\%$ of the target sensitivity at 4.5\arcsec and moderately exceed the target sensitivity at 48.6\arcsec.
    }
    \label{fig:ic10_beam_noise}
\end{figure*}

We test whether the LGLBS observations reach the project's proposed resolution and sensitivity goals using observations of the central field of IC10. This field was observed to the nominal depth using all four VLA configurations. After calibration, we pre-process the data, including continuum subtraction, following the steps outlined in \S\ref{sec:lglbs_hi_cd} using the PHANGS--ALMA imaging pipeline \citep{LEROY21_PIPELINE}. Specifically, we subtract the continuum component from the visibilities using CASA's \texttt{uvcontsub} task with a linear fit.
We fit the continuum component excluding the target's velocity range to avoid all line emission from the fit.
We also exclude the outer 10\% of the bandwidth (including the edge channels flagged by the pipeline) as we found that the continuum fits were biased to moderately lower amplitudes when including this portion of the bandwidth; this led to minor variation only for the brightest ($>50$~mJy) sources.
After continuum subtraction, we image a single $1.2$~\kms channel without detected \hi\ using different combinations of VLA array configurations and varying the weighting scheme used to grid the visibilities from uniform to robust (Briggs parameter from $r=-2$ to $2$) to natural. Figure \ref{fig:ic10_beam_noise} shows the rms intensity in the channel, the equivalent column density sensitivity over a 10~km~s$^{-1}$ window, and the \hi\ mass per beam for each imaging exercise. We show these for each array configuration and combination of array configurations and as a function of the beam semi-major axis.

During this imaging process, we noticed residual continuum emission from the brightest sources ($>100$~mJy) within the frequency range the we excluded from automated RFI flagging with the VLA pipeline.
In these cases, all emission (line and continuum) had a step-like function constant amplitude increase within the ``protected'' frequency range that RFI flagging was not applied on.
We suspect that this amplitude difference results from low-level RFI that contributed to the continuum component relative to the spectral line and is only statistically significant towards the brightest continuum sources.
Indeed, comparing against previous imaging from \citet{KOCH18,Koch2021} shows there is no significant offset in the total \hi\ emission over the whole galaxies.
To correct for this effect, we fit linear functions to the continuum-only channels, using aggressive sigma-clipping to exclude all \hi\ emission, within the frequency/velocity range where RFI-flagging was not applied and subtracted this model from the velocity range (including \hi\ emission) in these spectra.
This process removes all significant residual continuum emission, which we found was only present for a handful of bright sources in M31 and M33.

The figure shows that our data reach our target resolution and sensitivity, which are indicated by the labeled shaded gray lines in the center panel. This visualization also shows the benefits of combining the intermediate VLA configurations with the extended (A) and compact (D) baseline coverage. Our adopted 1:1:1:0.5 time-per-configuration ratio for A:B:C:D, combined with varying the weighting scheme yields a nearly continuous range of beam sizes with the target column density sensitivity from $2\mbox{--}60\arcsec$.

\section{LGLBS's 120 pc Resolution View of 21-cm \hi\ Emission}
\label{sec:lglbs_hi_cd}

\begin{figure*}[!t]
    \centering
    \includegraphics[width=1.0\textwidth]{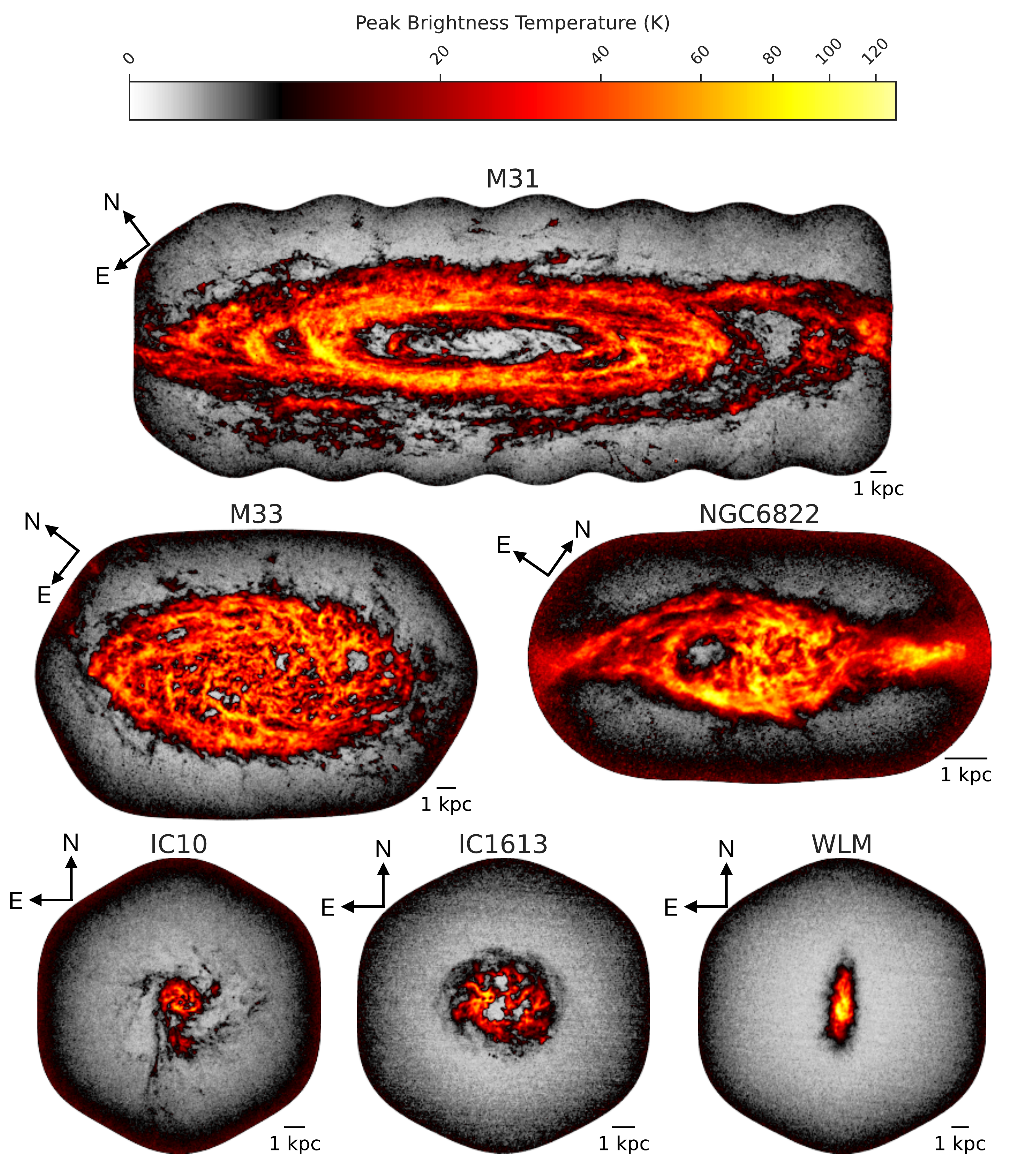}
    \caption{\textbf{\hi peak brightness temperature maps} for combined CD and GBT imaging over the full LGLBS field of view for each target. The images show the extent of emission and complex morphology visible in the bright \hi\ at high $\sim 100$~pc resolution.}
    \label{fig:hi_dwarfs_tpeak}
\end{figure*}

\begin{figure*}[!t]
    \centering
    \includegraphics[width=1.0\textwidth]{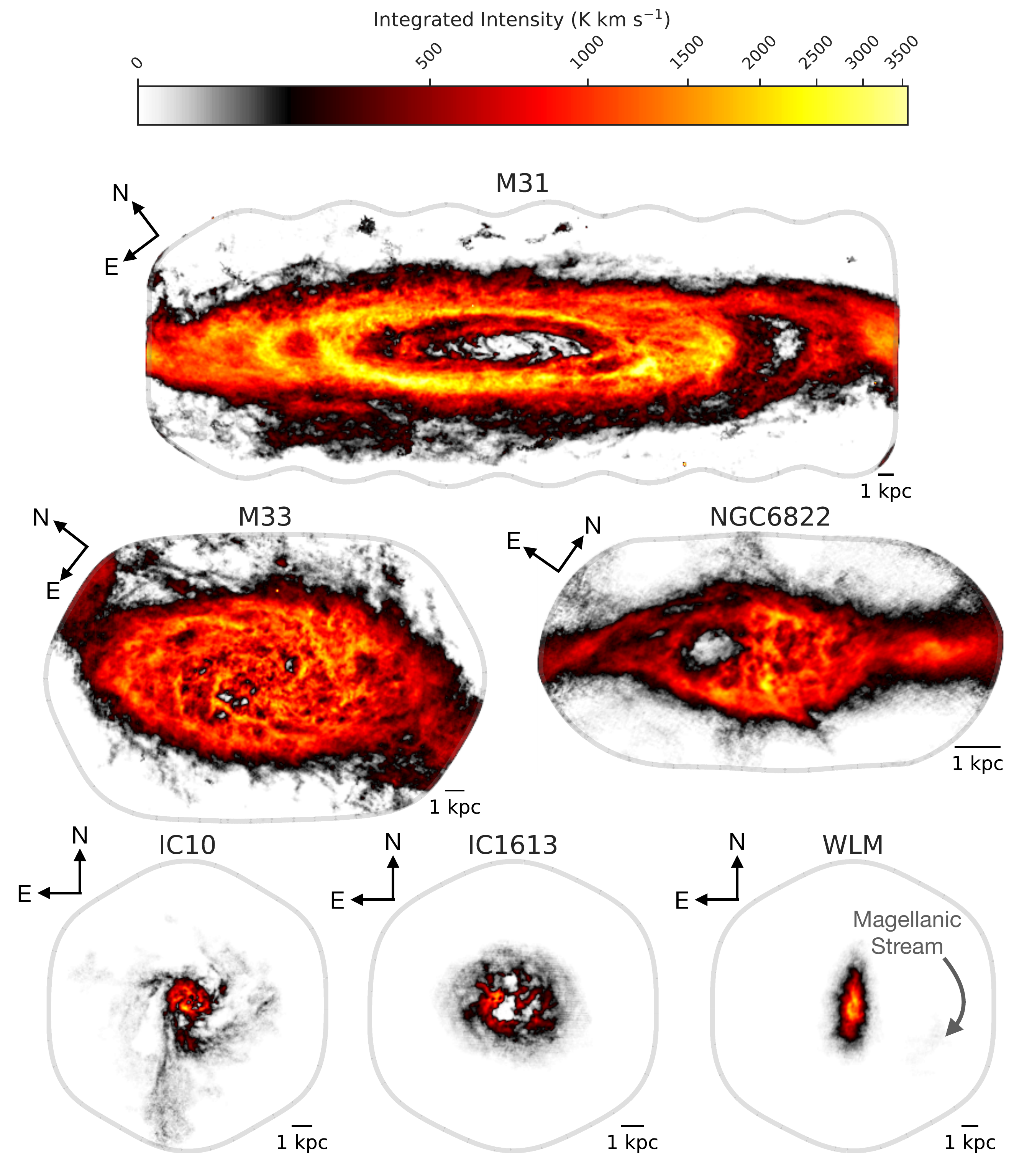}
    \caption{\textbf{\hi integrated intensity maps} of the four LGLBS dwarf galaxy targets at their native angular resolution. These maps show the wide-field extent combining CD configuration and GBT observations. The morphology of the \hi\ varies between these four galaxies, and we highlight in WLM where we recover faint \hi\ emission from the edge of the foreground Magellanic Stream \citep{Putman2003}.}
    \label{fig:hi_dwarfs_moment0}
\end{figure*}

\begin{figure*}[!t]
    \centering
    \includegraphics[width=0.9\textwidth]{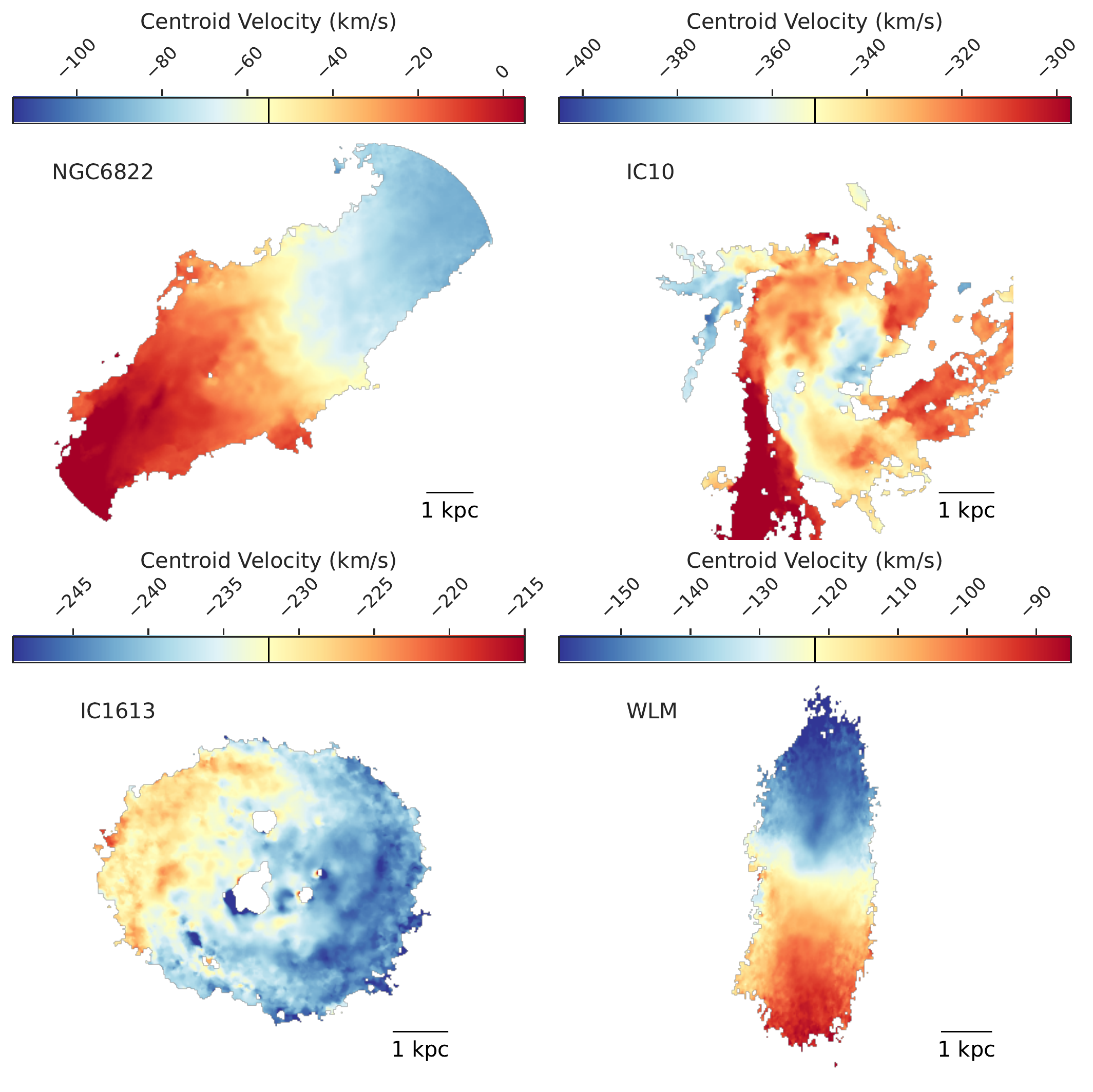}
    \caption{\textbf{\hi centroid velocity maps of the four LGLBS dwarf galaxies} at a common physical resolution of 120~pc. The colormaps are centered at the galaxy's systemic velocity indicated by the vertical black line in each colorbar and extend to $\pm V_{\rm max}$ (the maximum rotation velocity). For IC10, we show the velocity extent to $\pm 1.5 V_{\rm max}$ to better include the velocity ranges of the \hi\ tidal tails.}
    \label{fig:hi_dwarfs_mom1}
\end{figure*}

\begin{figure*}[!t]
    \centering
    \includegraphics[width=0.9\textwidth]{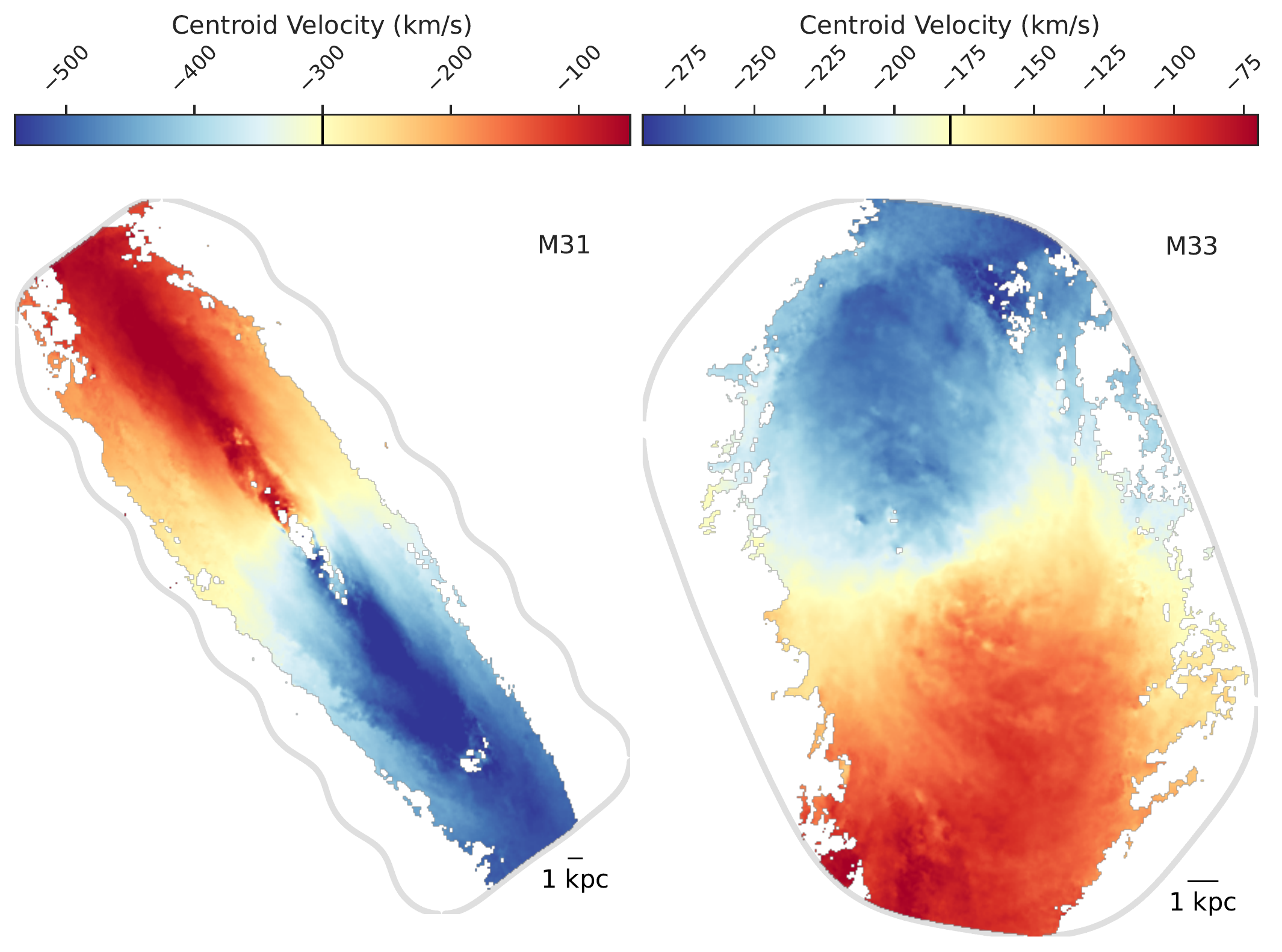}
    \caption{\textbf{\hi centroid velocity maps of M31 and M33} at a common physical resolution of 120~pc. The colormaps are centered at the galaxy's systemic velocity indicated by the vertical black line in each colorbar and extend to $\pm V_{\rm max}$ (the maximum rotation velocity).}
    \label{fig:hi_m31m33_mom1}
\end{figure*}

\begin{figure*}[!t]
    \centering
    \includegraphics[width=0.9\textwidth]{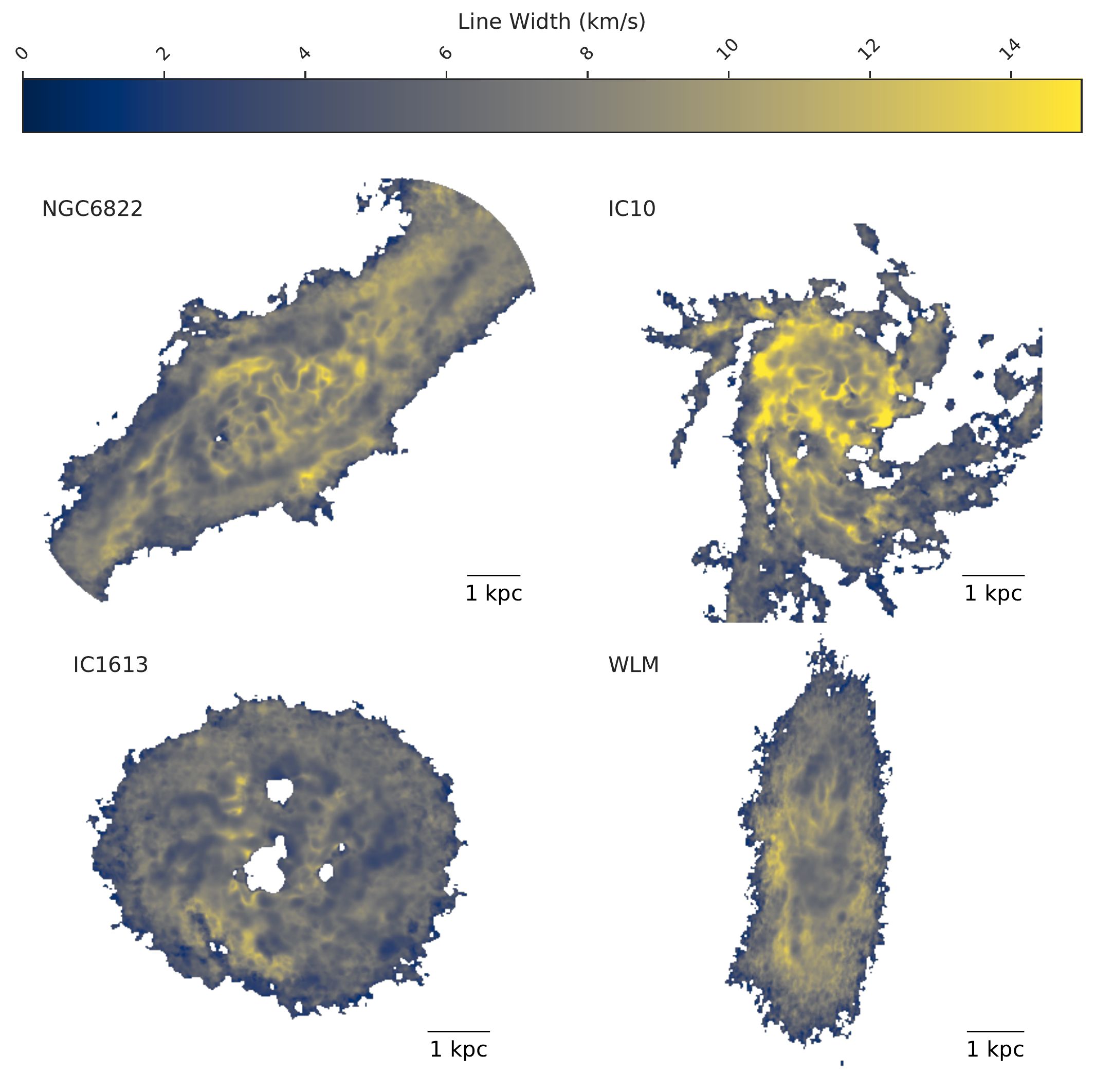}
    \caption{\textbf{\hi line width maps of the four LGLBS dwarf galaxies, as measured with the effective width,} defined as $\sigma = I_{\rm HI} / \sqrt{2\pi} \, T_{\rm peak}$. Each map is shown at a common physical resolution of 120~pc. Regions with recent massive star formation in NGC6822 and IC10 tend to show larger velocity dispersions, while IC1613 and WLM with lower recent star formation rates tend to have lower dispersions.}
    \label{fig:hi_dwarfs_ew}
\end{figure*}

\begin{figure*}[!t]
    \centering
    \includegraphics[width=0.9\textwidth]{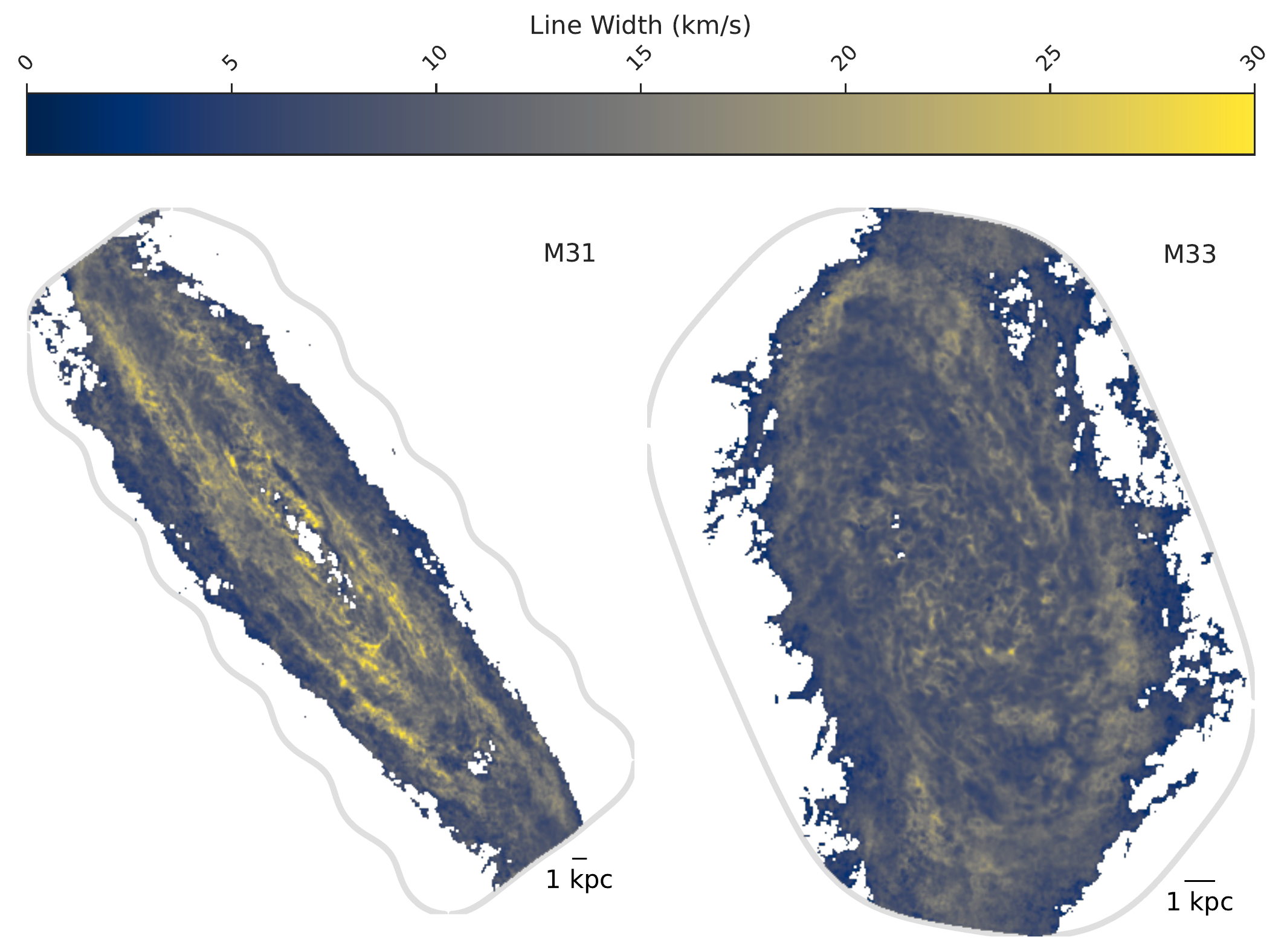}
    \caption{\textbf{\hi line width maps of M31 and M33, as measured with the effective width,} defined as $\sigma = I_{\rm HI} / \sqrt{2\pi} \, T_{\rm peak}$. Each map is shown at a common physical resolution of 120~pc. Both M31 and M33 have regions of larger line widths relative to the dwarf galaxies (Figure \ref{fig:hi_dwarfs_ew}), with the largest line widths in M31 likely resulting from multiple components along single sightlines from its high inclination angle \citep[e.g.,][]{Braun2009,Chemin2009,Koch2021}.}
    \label{fig:hi_m31m33_ew}
\end{figure*}

Figures \ref{fig:hi_dwarfs_tpeak}, \ref{fig:hi_dwarfs_moment0}, \ref{fig:hi_dwarfs_mom1}, \ref{fig:hi_m31m33_mom1}, \ref{fig:hi_dwarfs_ew}, and \ref{fig:hi_m31m33_ew} show the wide-field, $\sim20\arcsec$ resolution \hi\ images of the LGLBS targets. As described in \S\ref{sec:survey}, we observed seven pointing hexagonal mosaics in the C and D configurations for IC10, IC1613, and WLM. In NGC6822, we observed a common set of three pointings in a linear mosaic in all four configurations. We targeted M33 using a 13-point mosaic that spans the star-forming disk, and we cover the majority of M31 with a 49-point hexagonal mosaic. These images show the C and D configuration data imaged with robust $0.5$ weighting (as defined in CASA's \texttt{tclean}) and short spacing corrected using the GBT single dish data. Imaging using all four configurations will appear in N. Pingel et al. (in preparation).

We image the 21-cm emission for each target using an adapted version of the PHANGS--ALMA imaging pipeline \citep{Leroy2021_pipe}. We construct versions of the cubes with channel widths of 0.8, 2.1, and 4.1~\kms. We use a wide range of scales, 0, 10, 30, 100 and 300\arcsec , for the multi-scale deconvolution. Including these large scales is critical to achieve good results for the 21-cm emission, in which much of the flux often lies in an extended, relatively smooth distribution. After the multi-scale clean, we identify regions of significant emission and apply a single-scale clean down to a depth of $1$--$2\sigma$, considering the results to be converged when successive calls to \texttt{tclean} with a large number of components result in little change in the total flux of the clean model. The procedure for the PHANGS--ALMA imaging process resembles the one used to image VLA 21-cm emission from M33 \hi\ in \citet{Koch2018}, and so has already been shown to yield excellent results for a similar Local Group \hi\ data set. 

We feather the deconvolved VLA \hi\ cubes with the GBT \hi\ cubes (see Appendix \ref{app:gbt_data}) using \texttt{uvcombine} and the relative flux scaling factors measured in Appendix \ref{app:uvcombine}. For NGC 6822, Appendix \ref{app:galactic_hi_ngc6822} describes our process to create a NGC 6822-only GBT \hi\ cube with local pixelwise infilling \citep{Saydjari2022} and manual masking of Galactic \hi\ in the VLA cube.

Table \ref{tab:hi_masses} provides the $5\sigma$ \hi\ column density and mass per beam in all targets after convolving to a common physical scale of $120$~pc. We estimate the noise levels for each target from the rms scatter in 10 signal-free channels, considering only the central $120\arcmin$ of each mosaic, where the maps all have uniform sensitivity. Our \hi\ maps trace column densities of $1.4\mbox{--}4.1\times10^{19}$~cm$^{-2}$, equivalent to \hi\ surface densities of $0.1\mbox{--}0.3$~\msolpcsq, which extends far into the atomic-dominated regime as predicted by photodissociation region (PDR) models \citep[e.g.,][]{Krumholz2011} and well beyond the traditionally defined edge of the \hi\ disk at $\Sigma_{\rm HI} = 1$~\msolpcsq\ \citep[e.g.,][]{Wang2016}. Expressed as point mass sensitivity, these maps are sensitive to clouds of mass $1.8{-}4.1 \times 10^3$~M$_\odot$ per beam and thus able to detect even individual low mass clouds, including high velocity clouds \citep[e.g.,][]{putman2012} or the atomic gas associated with molecular clouds. 

We produce signal masks\footnote{We list further details and relevant parameters used for the masking with the data release.} and moment maps, again following the PHANGS--ALMA pipeline procedure \citet{Leroy2021_pipe}. In Figures \ref{fig:hi_dwarfs_tpeak}, \ref{fig:hi_dwarfs_moment0}, \ref{fig:hi_dwarfs_mom1}, \ref{fig:hi_m31m33_mom1}, \ref{fig:hi_dwarfs_ew}, and \ref{fig:hi_m31m33_ew}, we show the resulting integrated intensity, peak brightness temperature, centroid velocity, and effective width maps\footnote{Effective width is defined as $\sigma = I_{\rm HI} / \sqrt{2\pi} \, T_{\rm peak}$, where $I_{\rm HI}$ is integrated intensity and $T_{\rm peak}$ is the peak brightness temperature \citep[e.g.,][]{Sun2018}.}. These are the deepest $\sim 100$~pc wide-field views of \hi\ in these Local Group galaxies, and the figures show the excellent imaging quality of the VLA that we achieve with the LGLBS observing strategy.

We measure the total \hi\ masses in our fields of view from the integrated intensity maps produced using the ``broad'' signal mask \citep[see][]{Leroy2021_pipe}, which emphasizes flux completeness at the expense of moderately increasing the noise level. We adopt the sum of the integrated intensity error maps to estimate the uncertainty, corrected for the number of independent beam areas. Table \ref{tab:hi_masses} reports the measured atomic gas masses for these targets, assuming optically-thin \hi\ emission  (equivalent to a conversion factor of $0.0146$~\msolpcsq / K \kms when converting from \hi\ integrated intensity). Assuming that the absolute flux calibration has an inherent 5\% uncertainty, we also report an estimated systematic uncertainty which is dominated by the $\approx 5\%$ flux calibration uncertainty. Table \ref{tab:hi_masses} provides measurements for the \hi\ mass of each galaxy within our field. Based on the spatial extent of our targets, we expect that the area of our mosaic leads to a modest underestimate of the total \hi\ mass in M31, M33, and NGC6822.

In M31, within our field-of-view, we measured a total \hi\ mass of $5.7\times10^9$~\msol. This is in good agreement with the approximate $5 \times 10^9$~\msol\ quoted by \citet{Braun2004}, \citet{Yin2009} and close to the total \hi\ mass of $5.3 \times 10^9$~\msol\ estimated by \citet{Dame1993}, but larger by $30\%$ to the $4.1\times10^9$~\msol measured by \citet{Chemin2009}.
Our M31 map covers most of the disk but clearly misses some of the extended disk \citep[see][]{Braun2004,Braun2009}.

For M33, we find an \hi\ mass of $2.0\times10^9$~\msol, within 3\% of the \hi\ mass from \citet{Putman2009}. While some \hi\ recovered in the \citet{Putman2009} M33 map is excluded by our more limited field-of-view, we note that \citet{Putman2009} find a total mass of $3.6\times10^{8}$~\msol beyond the bright \hi\ disk (defined as below an \hi\ column density of $3.2\times10^{20}$~cm$^{-2}$), of which we recover $\approx30\%$ in the LGLBS map. Thus, the discrepancy for the missing \hi\ component falls within the total statistical and systematic uncertainty for our \hi\ mass measurement. Relative to \citet{KOCH18}, who calculate an \hi\ mass of $1.4\times10^9$~\msol using a portion of the C configuration VLA data and the same GBT data we use here, the \hi\ mass we find is $30\%$ larger. Similar to the larger flux correction factor we find (see Appendix \ref{app:uvcombine}), this difference reflects the improved VLA sensitivity and vastly improved short baseline coverage from including VLA D configuration data.

For NGC6822, our \hi\ mass measurement of $(1.9\pm0.1)\times10^8$ \msol\ is $20\%$ larger than the mass from \citet{deBlok2006} and 19\% larger than \citet{Namumba2017}. We expect our larger value is primarily due to differences in separating the Galactic \hi\ foreground, which we have carefully modeled and removed here (see Appendix \ref{app:galactic_hi_ngc6822}). This appears to represent a larger effect than the modest fraction of the extended disk missed by our mosaic.

The other three dwarf galaxies all lie completely within our fields of view with little concern regarding Milky Way contamination. For IC10, we find an \hi\ mass of $(7.5\pm0.1)\times10^7$~\msol\ that agrees within the uncertainty with the value from LITTLE THINGS \citep{Hunter2012} but is $14\%$ smaller than measurements from the GBT \citep{Nidever2013} and DRAO Synthesis Telescopes \citep{Namumba2019}. For IC1613, our \hi\ mass of $(6.3\pm0.1)\times10^7$~\msol\ is $59\%$ larger than the LITTLE THINGS value \citep{Hunter2012}, likely due to our inclusion of short-spacing data. Finally, in WLM, our \hi\ mass of $(7.4\pm0.1)\times10^7$~\msol\ agrees within the uncertainty with LITTLE THINGS \citep{Hunter2012} and is $9\%$ larger than previous VLA measurements from \citet{Kepley2007}. However, our VLA$+$GBT mass---similar to previous VLA measurements---is $31\%$ smaller than the GBT flux measurements from \citet{Hunter2011} and $11\%$ smaller than the GBT-only \hi\ mass we measure here.
Though these single dish \hi\ masses are themselves discrepant, the consistently higher single-dish \hi\ masses suggest a low-column density \hi\ component that falls below the VLA sensitivity.
Most of the previous VLA-based estimates tend to agree with our values, with the only exception being IC1613 where the short-spacing correction drives the difference.

We note that some of our \hi\ mass measurements differ from previous single-dish measurements despite the short-spacing corrections that we apply. Given our careful treatment for relative flux uncertainties prior to short-spacing correction (Appendix \ref{app:uvcombine}) and our uniform calibration of the archival GBT observations (Appendix \ref{app:gbt_data}), we are confident that the flux rescaling factors applied to the GBT cubes reflect our best total flux estimates when tied to the VLA absolute flux calibration. Therefore, we expect that the discrepant fluxes arise from differences in the spatial and spectral area used. 

\begin{deluxetable*}{lccccc}[t!]
\tabletypesize{\footnotesize}
\tablecaption{\label{tab:hi_masses} 21-cm \hi\ Imaging at 120~pc Resolution and \hi\ masses from combined VLA$+$GBT and GBT-only.}
\tablewidth{0pt}
\tablehead{
\colhead{Galaxy} & 
\colhead{Synthesized Beam} &
\colhead{$M_{\rm HI}$ (VLA$+$GBT) (\msol)} &
\colhead{$M_{\rm HI}$ (GBT-only)$^\dagger$ (\msol)} &
\colhead{$5\sigma$ \hi\ column density\tablenotemark{a}} & 
\colhead{$5\sigma$ \hi\ mass per beam\tablenotemark{a}} \\
\colhead{} & 
\colhead{(\arcsec)} &
\colhead{(\msol)} & 
\colhead{(\msol)} & 
\colhead{(cm$^{-2}$)} & 
\colhead{(\msol)} 
}
\startdata
M31$^\star$     & 32 & $(5.7\pm0.1\pm0.3)\times10^9$ & $(5.6\pm0.1\pm0.3)\times10^9$ & $3.1\times10^{19}$ & $4.1\times10^{3}$ \\
M33$^\star$     & 28 & $(2.0\pm0.1\pm0.1)\times10^9$ & $(2.0\pm0.1\pm0.1)\times10^9$ & $2.2\times10^{19}$ & $2.9\times10^{3}$ \\
IC10            & 31 & $(7.5\pm0.1\pm0.4)\times10^7$ & $(7.7\pm0.1\pm0.3)\times10^7$ & $2.3\times10^{19}$ & $2.9\times10^{3}$ \\
IC1613          & 33 & $(6.3\pm0.1\pm0.3)\times10^7$ & $(6.3\pm0.1\pm0.3)\times10^7$ & $1.8\times10^{19}$ & $2.3\times10^{3}$ \\
NGC6822$^\star$ & 48 & $(1.9\pm0.1\pm0.1)\times10^8$ & $(1.8\pm0.1\pm0.1)\times10^8$ & $1.4\times10^{19}$ & $1.8\times10^{3}$ \\
WLM             & 27 & $(7.4\pm0.1\pm0.4)\times10^7$ & $(8.2\pm0.1\pm0.4)\times10^7$ & $4.1\times10^{19}$ & $3.9\times10^{3}$ 
\enddata
\tablenotetext{a}{The $5\sigma$ \hi\ column density and mass sensitivities are measured over 10~\kms.}
\tablenotetext{\star}{Our VLA$+$GBT mass measurements are limited by the LGLBS spatial coverage.}
\tablenotetext{\dagger}{GBT mass measurements are made at the native GBT resolution over a field-of-view that extends beyond the VLA coverage. These masses also incorporate the flux calibration corrections derived for feathering in Appendix \ref{app:uvcombine}.}
\tablecomments{Total \hi\ mass, $5\sigma$ column density, and mass per beam sensitivities measured on matched 120~pc scales from CD configuration imaging combined with GBT short spacing data. The two uncertainties for the mass values correspond to the statistical and systematic uncertainty, respectively, where we assume a systematic uncertainty of 5\% from flux uncertainty.}
\end{deluxetable*}

Over this small sample size of four dwarf galaxies, we see large variations in the gas morphology. IC10 shows clear, $>$kpc-long extended \hi\ structures whose morphology resembles tidal tails caused by a recent interaction \citep{Nidever2013}. NGC6822 also shows an extended \hi\ component relative to the compact stellar disk, which previous work has also suggested may be evidence of a tidal tail \citep{deBlok2006}. In IC1613, we do not detect extended \hi\ emission above our column density sensitivity ($\sim2\mbox{--}4\times10^{19}$~cm$^{-2}$) beyond the galaxy disk. In WLM, a faint extended feature marked in Figure~\ref{fig:hi_dwarfs_moment0} corresponds to \hi\ foreground emission associated with the Magellanic Stream \citep{Putman2003}, and we see weak filamentary structures extending beyond the disk (Caballero Vargas et al., in prep.). In both IC1613 and WLM, the observed morphology mostly shows a rotating, inclined HI disk.

The \hi\ kinematics (Figs.~\ref{fig:hi_dwarfs_mom1} and \ref{fig:hi_dwarfs_ew}) also show diversity and complexity. Visually, the inner disks of NGC6822, IC10, and WLM are relatively symmetric about the systemic velocity. This is expected for motions dominated by circular rotation and agrees well with previous \hi\ kinematic modeling of these sources \citep{Oh2015,Namumba2017,Namumba2019,Ianjamasimanana2020}. On the other hand, IC1613 appears to have an asymmetric velocity field about $V_{\rm sys}$. 

Meanwhile, the centroid velocities in M31 and M33, shown in Figure \ref{fig:hi_m31m33_mom1}, exhibit a typical circular-rotation pattern on large scales. Future LGLBS work will explore fine-scale kinematic variations and non-circular motions, leveraging our high angular and velocity resolution and building on the large body of literature dedicated to studying M31 and M33 \citep[including][and references in Table \ref{tab:lit}]{Corbelli1997,Corbelli2000,Corbelli2010,Kam2017,Koch2018}. Figure \ref{fig:hi_m31m33_mom1} also shows resolved \hi\ detected well-beyond the bright \hi\ disks to the edges of the LGLBS's fields of view, where the \hi\ kinematics offer potential new insight into the neutral gas being accreted or ejected. These extended ``clouds'' of \hi\ have been studied in both galaxies \citep[e.g.,][]{Thilker2004,Westmeier2005,Putman2009,Keenan2016,Koch2018}, but, for most of them  LGLBS will yield the highest resolution and most sensitive view to date, and often the first interferometric imaging.

In Fig. \ref{fig:hi_dwarfs_ew}, NGC6822 and IC10 show larger line widths in their inner disks ($>10$~\kms) compared to the other two targets. This likely reflects the combined effect of the disturbed \hi\ environment from potential recent interactions and turbulent driving from massive stellar feedback. Both of these galaxies host massive star forming regions, with IC10 currently under-going a localized starburst \citep{Hunter2001}. Reflecting their morphology, ordered large-scale kinematics, and lower star formation rates, IC1613 and WLM have lower line widths, though they do show localized enhancements in the vicinity of regions with known recent star formation. WLM's velocity dispersion is moderately larger than IC1613's, which is expected due to our nearly edge-on view of WLM \citep{Ianjamasimanana2020}.

Figure \ref{fig:hi_m31m33_ew} shows that M31 and M33 have regions with larger velocity dispersions ($>15$~\kms) than what we find in the four dwarf galaxies, consistent with previous work \citep{Koch2018,Utomo2019,Koch2021}. These larger velocity dispersions result from several sources, including stellar feedback from wide-spread massive star formation that can drive elevated levels of turbulence \citep[e.g.,][]{MacLow2004} 
and, particularly for M31, complex sightlines due to where multiple clouds are superimposed along the line of sight due to the high inclination angle. Using pilot LGLBS observations of M31 and M33, \citet{Koch2018} and \citet{Koch2021} demonstrated that high velocity resolution and sensitivity allow one to separate distinct kinematic features along individual lines of sight in these galaxies. A key focus of upcoming LGLBS work will be to conduct kinematics analyses that separate individual components to disentangle the different physical sources driving the complex \hi\ line shapes.

These images provide a preview of the richness of the LGLBS \hi\ maps.  In terms of spatial resolution and sensitivity, LGLBS is a clear precursor to the \hi\ mapping capabilities that the SKA \citep{SKA} and ngVLA \citep{Murphy2018} will provide in the coming decades.

\section{Summary and Outlook}
\label{sec:summary}

We present the Local Group L-band Survey (LGLBS), a Karl G. Jansky Very Large Array (VLA) program to image 21-cm \ion{H}{1} and OH line emission and radio continuum from the six VLA-visible Local Group (all $D < 1$ Mpc) galaxies that show evidence for recent massive star formation and have \hi\ mass $\log_{10} (M_{\rm HI}/{\rm M}_\odot)> 7$. By focusing on these closest targets and using all four VLA configurations, we achieve an unprecedented combination of physical resolution and sensitivity. This paper presents the survey design, describes our calibration and quality assurance procedures, and shows some first results for imaging using the C and D configurations. Key points include:

\begin{enumerate}
\item The survey targets all \hi -rich Local Group galaxies visible from the VLA site: the massive spiral M31 (Andromeda), the dwarf spiral M33 (Triangulum), and the four dwarf irregulars IC10, IC1613, NGC6822, and WLM. Our survey complements the ongoing GASKAP-HI survey \citep{Pingel2022}, which will produce similar quality data for the LMC and SMC.

\item We take advantage of the VLA's powerful WIDAR correlator to simultaneously observe the 21-cm line at sharp 0.4~km~s$^{-1}$ velocity resolution, the full 1--2 GHz polarized continuum, and the OH lines at 1612, 1665, 1667, and 1720 MHz (Fig. \ref{fig:lband_tunin}). We also record information on radio recombination lines and enabled simultaneous REALFAST observations by another team to search for transients such as fast radio bursts \citep[][]{LAW2018-realfast}. The 21-cm observations trace emission and absorption from atomic gas, which makes up most of the interstellar medium in each of our targets. The continuum observations mostly trace a mixture of free-free emission associated with \hii\ regions, synchrotron emission powered by current and past supernova remnants, and background sources. The OH main lines (1667 MHz, 1665 MHz) provide a faint but potentially powerful probe of the molecular ISM that complements CO and dust emission \citep{Busch2024}, while the OH satellite lines (1612 MHz, 1720 MHz) can exhibit strong masers from the ISM which probe physical conditions of shocked gas or stellar activity \citep{Lockett1999,Gray2005,KochMaser2018}.

\item Our observations cover the area of high column density \hi\ and star formation activity in each target, and we observe each target using each of the VLA's four configurations (A, B, C, and D). This combination yields 21-cm maps with physical resolutions as high as $10{-}20$~pc and continuum maps with resolution as high as $5{-}10$~pc. Combined with our $0.4$~km~s$^{-1}$ channel width, LGLBS achieves a detailed radio view of the atomic gas, \hii\ regions, and supernova remnants previously only accessible in the Milky Way. 

\item As of this publication, all observations have concluded and $>430$ tracks ($\gtrsim 1,800$ h of observations) have been calibrated and manually quality assured. We present a new software framework that enables this QA. The framework deploys a slightly modified version of the VLA pipeline to calibrate each track and then produces a series of interactive plots that allow in depth inspection of the visibility data using a browser-based interface. This framework allowed our distributed team to efficiently QA a large amount of data. The approach is scalable and can be adapted for use in other projects dealing with widely distributed collaborations and large data volumes and is readily applicable to non-LGLBS VLA L-band observations taken with the WIDAR correlator.
We make our code publicly available via the \texttt{ReductionPipeline}\footnote{\url{github.com/LocalGroup-VLALegacy/ReductionPipeline}} and \texttt{QAPlotter}\footnote{\url{github.com/LocalGroup-VLALegacy/QAPlotter}} packages.

\item We present first 21-cm images for our six targets using data from the compact C+D configurations, combined with GBT short spacing data. These show good overall agreement with previous large scale measurements, reveal extended atomic gas distributions down to column densities of $\approx 2{-}4 \times 10^{19}$~cm$^{-2}$ ($5\sigma$), and $5\sigma$ sensitivity to individual \hi\ clouds with masses as low as $\gtrsim 2{-}5 \times 10^3$~M$_\odot$.

\end{enumerate}

Even after calibration and quality assurance, imaging the data represents a significant technical challenge. The 21-cm imaging for our combined ABCD data is described in detail by N. Pingel et al. (in preparation) \citep[and see][]{Pingel2024}, while the Stokes $I$ continuum imaging is described by S. Sarbadhicary et al. (in preparation). All images and data cubes from LGLBS will be publicly available, with the CD+GBT data presented in this paper publicly released here\footnote{\textbf{\url{https://www.canfar.net/storage/vault/list/LGLBS/RELEASES/LGLBS-HI-v1.0}; to be updated with DOI from CANFAR.}}, and initial data releases of the high resolution line and continuum data will accompany the continuum and line imaging papers.

\begin{acknowledgments}
We gratefully acknowledge the anonymous reviewer for their comments that improved the clarity of this paper.
We are grateful for the support of the staff at the National Radio Astronomy Observatory who facilitated this ``Extra Large'' program.  Our survey design benefited from an extensive technical review led by Claire Chandler, and we are grateful for the work of the VLA scheduling team, especially Amy Mioduszewski, for helping facilitate our observational requirements.
We thank Filippo Maccagni for sharing his compilation of archival \hi\ surveys that we reproduce in Figure \ref{fig:lglbs_to_hi_surveys}.
The National Radio Astronomy Observatory is a facility of the National Science Foundation operated under cooperative agreement by Associated Universities, Inc.
This research was enabled in part by support provided by the Digital Research Alliance of Canada (\url{alliancecan.ca}). 
This research used the Canadian Advanced Network For Astronomy Research (CANFAR) operated in partnership by the Canadian Astronomy Data Centre and The Digital Research Alliance of Canada with support from the National Research Council of Canada the Canadian Space Agency, CANARIE and the Canadian Foundation for Innovation.
Some of the computations in this paper were conducted on the Smithsonian High Performance Cluster (SI/HPC), Smithsonian Institution (\url{https://doi.org/10.25572/SIHPC}).
The LGLBS collaboration is grateful for support from the National Science Foundation, AST-2205628 and AST-205630.
EWK acknowledges support from the Smithsonian Institution as a Submillimeter Array (SMA) Fellow and the Natural Sciences and Engineering Research Council of Canada.
LC also acknowledges support from NSF grant AST-2107070.
ER and HC acknowledge the support of the Natural Sciences and Engineering Research Council of Canada (NSERC), funding reference number RGPIN-2022-03499.
TW gratefully acknowledges support from the Chu Family Foundation for the Vermilion River Fund for Astronomical Research at the University of Illinois.
SS gratefully acknowledges support provided by the University of Wisconsin - Madison
Office of the Vice Chancellor for Research and Graduate Education with funding from the Wisconsin Alumni Research Foundation.
DJP is supported through the South African Research Chairs Initiative of the Department of Science and
Technology and National Research Foundation (Grant number 77825).
JS acknowledges support by the National Aeronautics and Space Administration (NASA) through the NASA Hubble Fellowship grant HST-HF2-51544 awarded by the Space Telescope Science Institute (STScI), which is operated by the Association of Universities for Research in Astronomy, Inc., under contract NAS~5-26555.

\end{acknowledgments}

\vspace{5mm}
\facilities{VLA, GBT}

\software{astropy \citep{ASTROPY13,ASTROPY18,ASTROPY22},
plotly, 
radio-astro-tools (spectral-cube, radio-beam, uvcombine) \citep{SPECTRALCUBE2020}, 
CASA \citep{CASA}, 
GBTIDL \citep{Marganian2013}, 
seaborn \citep{seaborn}, 
matplotlib \citep{mpl}, 
numpy \& scipy \citep{2020SciPy-NMeth}, 
PHANGS imaging pipeline \citep{LEROY21_PIPELINE}.}

\appendix

\section{GBT Calibration and Imaging}
\label{app:gbt_data}

The \hi\ emission in our targets is extended relative to both the VLA primary beam and the largest recoverable spatial scales in D configuration. Thus, we require short-spacing corrections with single dish observations to capture the large-scale \hi\ emission. For this purpose, we use archival GBT data from several observing projects. Most used the now-decommissioned GBT Spectrometer backend, while some later projects used the Versatile GBT Astronomical Spectrometer (VEGAS).

We re-reduced the GBT spectrometer data from Green Bank Observatory (GBO) projects AGBT12B\_312 (PI: Johnson, targets: WLM and IC10), AGBT13A\_430 (PI: Ashley, target: IC10), and AGBT13B\_169 (PI: Johnson, target: NGC6822). The observations, GBT spectrometer set up, and observing strategy are similar to what is described in \citet{Johnson2013}. In brief, the total bandwidth was 12.5 MHz with 16,384 channels, and the band centered on the frequency of the \hi\ line at the heliocentric redshifts of the sources. For calibration purposes, the band was frequency-switched $-$3.5 MHz from the center frequency with a one-second period. We re-reduce the spectrometer spectra using the \texttt{getfs} procedure in the GBTIDL\footnote{\url{https://gbtidl.nrao.edu/}} data reduction package. We then fit and subtract a second-order polynomial baseline across emission-free channels to account for any residual structure and remove continuum after the \texttt{getfs} procedure. The spectra are scaled assuming a constant zenith opacity of 0.01 to convert into units of opacity-corrected antenna temperature ($T_A^*$). These calibrated spectra are smoothed to a final resolution of 5.15 \kms using a Gaussian kernel.

The VEGAS set up and observing strategy for AGBT16A\_413 (PI: Pisano) targeting IC1613 is similar to that outlined in \citet{Pingel2018}. The total bandwidth was 23.44 MHz with 32,768 channels with the band centered on the \hi\ line at the heliocentric redshifts of the source. The band was frequency-switched by $\pm$5.5 MHz. Section 3.1 in \citet[][] {Pingel2018} presents the details on the flux calibration procedure, including the custom GBTIDL routines that build a reference spectrum from the first four and last four integrations at the edge of the 2$^\circ\times2^\circ$ on-the-fly maps. Additionally, they account for a crosstalk issue in the backend electronics that was present in the early VEGAS science data and affects the flux calibration. To convert to units of ($T_A^*$), we use the same scaling and baseline fitting procedure as described above for the GBT spectrometer data, and also smooth these data to a final spectral resolution of 5.15 \kms . 

Before imaging, we resample the spectral data onto a common grid using the \texttt{gshift} GBTIDL procedure. We also convert the Galactic coordinates in the headers of the spectrometer SDFITS files\footnote{\url{https://fits.gsfc.nasa.gov/registry/sdfits.html}} to J2000 equatorial coordinates. The data are then combined and imaged using the \texttt{ gbtgridder}\footnote{\url{https://github.com/GreenBankObservatory/gbtgridder}} tool. We adopt a gain of 1.86 K Jy$^{-1}$, based on detailed modeling of the aperture illumination of the GBT at L-Band, and assume the FWHM of the GBT L-Band beam to be 9.1$'$ \citep{Pingel2018,sardone2021}.

\begin{deluxetable}{lccc}[t]
\tabletypesize{\footnotesize}
\tablecaption{\label{tab:sdfluxratio} Fitted VLA-to-GBT flux scaling factors applied to the GBT \hi\ cubes prior to short-spacing correction with feathering. The velocity range indicates the spectral channels incorporated into the scaling factor measurement.}
\tablewidth{0pt}
\tablehead{
\colhead{Galaxy} & 
\colhead{GBT flux scaling factor} & 
\colhead{Velocity range (\kms)} & 
}
\startdata
M31     & 1.00 & $-500$ to $-100$  \\
M33     & 1.10 & $-260$ to $-100$ \\
NGC6822 & 0.92 & $-75$ to $-25$ \\
IC10    & 0.96 & $-380$ to $-260$  \\
IC1613  & 0.89 & $-265$ to $-210$ \\
WLM     & 1.14 & $-170$ to $-70$ \\
\enddata
\end{deluxetable}

\section{Relative Flux Calibration for \hi\ Short-Spacing Correction}
\label{app:uvcombine}

To ensure consistency when combining the interferometer and single dish data, we test whether the absolute flux calibrations agree within the angular scales measured by both data sets. For these tests we focus on VLA data in the C and D configurations, i.e., those with the shortest baseline coverage and largest overlap with GBT in angular scales sampled.

Prior to comparing the flux at common angular scales, we smooth (if necessary) and resample the GBT cubes to the $4.1$~\kms of our ``low spectral resolution'' VLA \hi\ cubes. We then check the spatial registration of the VLA to these velocity-matched GBT cubes with the \texttt{image\_registration} package \citep{image_registration}. We find spatial offsets consistent within $<20\%$ of the GBT beam size and therefore apply no spatial shifts to the GBT cubes. Finally, we reproject the GBT cubes to the same spatial pixel grid as the VLA data.

We use the \texttt{uvcombine} package \citep{uvcombine} and apply a similar set of tests as described in \citealt{Koch2018}, which are based on \citet{STANIMIROVIC99}. Using the reprojected GBT cubes, we sample the data on angular scales between 9.5$\arcmin$ and $18\arcmin$, corresponding to baseline lengths of $100$ to $40$~m, and aggregate the distribution of flux measurements over the \hi\ velocity range for each target. Following \citealt{Koch2018}, we model the natural log of the flux ratios with a Cauchy distribution, which naturally accounts for excess tails in the distribution that result from noise fluctuations in either the numerator or denominator of the ratio. We fit this model to the measured flux ratios using a maximum-likelihood approach. The location of the peak in the best-fit model corresponds to the interferometer-to-single dish flux scaling factor. 
Table \ref{tab:sdfluxratio} gives the fitted flux scaling factors and the velocity range to which the flux ratios are fit.

We note that the M33 flux ratio value of $1.10$ given in  Table \ref{tab:sdfluxratio} is larger than the value of $1.02$ from \citet{KOCH18}, despite using the same GBT data.
The gridded GBT data have a larger effective beam width of $9.8\arcmin$ as they were gridded with a Gaussian kernel, not the Gaussian-Bessel kernel, to account for the minimal spatial overlap between the 4 observed maps \citep[see further details in][]{LOCKMAN2012}.
We perform the same series of checks from \citet{KOCH18} and find no other systematic effects to explain this discrepancy in the flux ratio value.
After thorough checks, we find that this difference persists primarily from the inclusion of the LGLBS D configuration observations significantly increasing the short baseline coverage relative to \citet{KOCH18} that used only a subset of the C configuration data included here.
As including the D configuration is ideal for these $uv$-overlap tests, we adopt the new flux ratio of $1.10$ here.

Table \ref{tab:sdfluxratio} shows the fitted flux scaling factors, which we apply to the GBT data before feathering. Similar to previous work, we adopt the VLA flux calibration as the reference and apply correction factors to only the GBT data. The scaling factors in Table \ref{tab:sdfluxratio} tend to be within $10\%$ of unity, and thus appear broadly consistent with the expected systematic uncertainty in the flux calibration of both data sets. The values we find are typical from studies using similar combinations of VLA and GBT data with various other telescopes \citep[e.g.,][]{blagrave2017,KOCH18,KOCH21HI}.

Although we derive these flux scaling factors using only the compact VLA C and D configurations, we use them for all VLA+GBT combinations. The extended VLA configurations will be on a consistent \citet{Perley2017} flux scale with the C and D configurations.

\section{Modeling and Separating the Galactic \hi\ Foreground in NGC6822}
\label{app:galactic_hi_ngc6822}

NGC6822 has significant velocity overlap with Galactic \hi\ near recession velocity $0$~\kms. The foreground \hi\ is bright and extended relative to the VLA primary beam due to a foreground Galactic molecular cloud \citep{GRATIER10}. To separate the foreground \hi\ emission spatially overlapping with NGC6822, we developed a new approach that utilizes the local pixel-wise infilling (LPI) technique developed by \citet{SAYDJARI22} to allow photometry on structured backgrounds. 

First, we use the rotation curve model derived in \citet{NAMUMBA17} to spatially mask NGC6822 in the GBT cube. To extrapolate the rotation curve beyond radii included in their measurements, we fit an arctan function to their circular rotation data. We find that some \hi\ emission in these outer regions deviates from the single orientation (inclination/positions angle) of the NGC6822 rotation curve from \citet{NAMUMBA17} and is thus not included in the mask. To account for this deviation, we simply extend the mask, at all radii, spectrally by $\pm20$~\kms and spatially by  1 beam width using binary dilation. 
We do not apply this binary dilation operation for velocity channels between $-5$ to $20$~\kms where we visually confirm that all NGC6822 emission is contained within the original non-dilated mask. In those channels we do not apply any spatial extension of the mask.
We note that these kinematics deviations that are recovered in our resolved, deep \hi\ observations will be explored with updated circular rotation curve modeling in future work.  For our purposes here, however, it is sufficient to overly mask to ensure that the mask thoroughly excludes all \hi\ emission from NGC6822, even if some Galactic \hi\ emission is included in the mask.
We visually inspect the GBT cube after masking NGC6822 to ensure that only Galactic \hi\ emission remains.

We then build a covariance matrix from the remaining purely Galactic \hi\ emission and infill the masked region in each channel using LPI implemented in \texttt{CloudClean.jl}\footnote{\url{github.com/andrew-saydjari/CloudClean.jl}}. We build a local covariance matrix of size $129\times129$ and an infill region size of $51$ pixels, matching the largest masked area (in the NGC6822 GBT data cube one pixel has linear size $3.5\arcmin$). We infill the masked regions using the mean LPI distribution in each channel. We then subtract the infilled Galactic \hi\ cube from the original GBT cube to produce a NGC6822-only \hi\ cube. We use this cube for the feathering described above in \S\ref{app:uvcombine}, though note that the infill modeling does not affect the relative flux calibration measurement because we use channels without strong foreground \hi\ for the flux calibration comparison. 

\begin{figure*}[!t]
    \centering
    \includegraphics[width=1.0\textwidth]{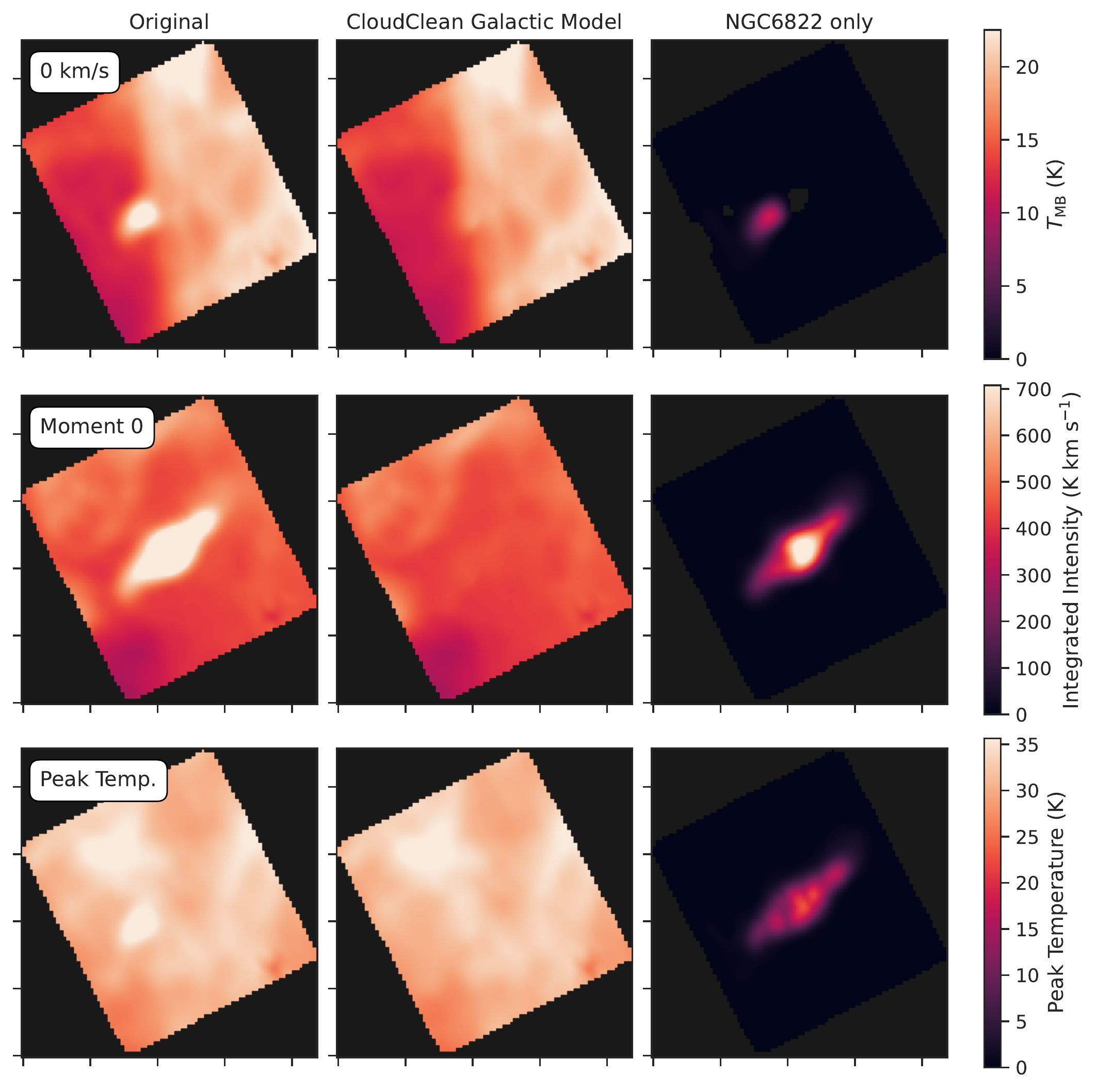}
    \caption{\textbf{Foreground Galactic \hi\ separation in NGC6822 using \texttt{CloudClean.jl}} \citep{SAYDJARI22}.
    We show the original GBT cube (left column), the Galactic \hi\ only model produced with \texttt{CloudClean.jl} (center column), and the NGC6822 only cube from subtracting the Galactic model from the original (right column).
    The rows show a single channel at recession velocity 0~\kms (top), integrated intensity (moment 0, middle), and peak temperature (bottom).
    LPI shows excellent results for separating Galactic \hi\ foreground from NGC6822 emission where both overlap in velocity.
    }
    \label{fig:ngc6822_foregroundHI}
\end{figure*}

Figure \ref{fig:ngc6822_foregroundHI} shows the infill results in one channel, the integrated intensity map, and the peak temperature map for the original data set, the Galactic \hi-only model from \texttt{CloudClean.jl}, and the final NGC6822-only cube. LPI produces an excellent foreground \hi\ model which, when subtracted from the original cube, leaves behind only the emission from NGC6822.

From the NGC6822-only cube, adopting a distance of $526\pm25$~kpc \citep[][; Table \ref{tab:physical}]{lee2024dist}, we measure a total luminosity of $(1.21\pm0.06)\times10^{10}$~K~\kms~pc$^2$ or an optically-thin \hi\ mass of $(1.8\pm0.1)\times10^8$~\msol, which includes the flux scaling correction factor of $0.92$ (see Appendix \ref{app:uvcombine}).
This \hi\ mass is $20\%$ larger than the total \hi\ mass of $1.5\times10^8$~\msol\ measured with Parkes observations by \citet[][adjusting for a distance of 520 kpc that we adopt here]{deBlok2006}. This small increase likely results from our pixelwise infilling approach to modeling the Galactic \hi, as the interpolation scheme used by \citet{deBlok2006} will produce an overly-smoothed version of the Galactic \hi\ foreground.

While most foreground Galactic \hi\ is filtered out by the VLA, some fine-scale extended \hi\ filaments that are clearly visible in the GBT map are still detected in the C+D cubes at recession velocities of $\sim-30\mbox{--}10$~\kms.  These are spatially separated from NGC6822's red-shifted limb and we manually mask this emission prior to calculating \hi\ properties and moment maps. The result of this combined LPI and manual masking leads to a data cube containing our best estimate of emission only from NGC 6822.

\bibliography{lglbs-master,akl_lglbs}{}

\bibliographystyle{aasjournal}

\end{document}